\documentclass[12pt]{article}
\usepackage{color}

\usepackage{graphicx}

\usepackage{amsfonts}
\usepackage{amsmath}
\usepackage{amssymb}
\usepackage{theorem}
\usepackage{booktabs}
\usepackage{rotating}
\usepackage[colorlinks,linkcolor=red,anchorcolor=blue,citecolor=blue]{hyperref}
\allowdisplaybreaks

\newtheorem{theorem}{Theorem}

\newtheorem{proposition}{Proposition}
\newtheorem{remark}{Remark}

\newtheorem{example}{Example}

\catcode`@=11 \@addtoreset{equation}{section} \catcode`@=12

\numberwithin{equation}{section}
\newcommand{\ba}{\begin{equation}}
\newcommand{\ea}{\end{equation}}

\newcommand{\0}{\mbox{\boldmath $0$}}
\newcommand{\fS}{\mbox{\boldmath $S$}}

\newcommand{\fz}{\mbox{\boldmath $z$}}

\newcommand{\mH}{\mbox{\textup{\textbf{H}}}}

\newcommand{\mU}{\mbox{\textup{\textbf{U}}}}
\newcommand{\mV}{\mbox{\textup{\textbf{V}}}}
\newcommand{\fgamma}{\mbox{\boldmath $\gamma$}}

\newcommand{\fTheta}{\mbox{\boldmath $\Theta$}}
\newcommand{\ftheta}{\mbox{\boldmath $\theta$}}
\newcommand{\fThetai}{\mbox{\scriptsize\boldmath $\Theta$}}
\newcommand{\fthetai}{\mbox{\scriptsize\boldmath $\theta$}}
\newcommand{\fvtheta}{\mbox{\boldmath $\vartheta$}}
\newcommand{\fvthetai}{\mbox{\scriptsize\boldmath $\vartheta$}}
\newcommand{\fnu}{\mbox{\boldmath $\nu$}}
\newcommand{\fnui}{\mbox{\scriptsize\boldmath $\nu$}}

\newcommand{\skellam}{\textup{Sk}}
\newcommand{\poi}{\textup{Poi}}
\newcommand{\bin}{\textup{Bin}}

\newcommand{\bbn}{\mathbb{N}}
\newcommand{\bbr}{\mathbb{R}}
\newcommand{\bbz}{\mathbb{Z}}

\newcommand{\iid}{i.\,i.\,d.}
\newcommand{\ie}{i.\,e., }
\newcommand{\eg}{e.\,g., }

\usepackage{dsfont}
\newcommand{\indfkt}{\mathds{1}}

\usepackage{natbib}

\usepackage{hyperref}

\begin{document}

\title{Tobit Models for Count Time Series}
\author{
Christian H.\ Wei\ss{}\thanks{
Helmut Schmidt University, Department of Mathematics and Statistics, Hamburg, Germany. E-Mail: \href{mailto:weissc@hsu-hh.de}{\nolinkurl{weissc@hsu-hh.de}}. ORCID: \href{https://orcid.org/0000-0001-8739-6631}{\nolinkurl{0000-0001-8739-6631}}.}
\and
Fukang Zhu\thanks{
School of Mathematics, Jilin University, 2699 Qianjin, Changchun 130012, China. E-Mail: \href{mailto:zfk8010@163.com}{\nolinkurl{zfk8010@163.com}}. ORCID: \href{https://orcid.org/0000-0002-8808-8179}{\nolinkurl{0000-0002-8808-8179}}.}\ \thanks{Corresponding author.}
}
\date{}
\maketitle
\begin{abstract}
\noindent
Several models for count time series have been developed during the last decades, often inspired by traditional autoregressive moving average (ARMA) models for real-valued time series, including integer-valued ARMA (INARMA) and integer-valued generalized autoregressive conditional heteroscedasticity (INGARCH) models. Both INARMA and INGARCH models exhibit an ARMA-like autocorrelation function (ACF). To achieve negative ACF values within the class of INGARCH models, log and softplus link functions are suggested in the literature, where the softplus approach leads to conditional linearity in good approximation. However, the softplus approach is limited to the INGARCH family for unbounded counts, \ie it can neither be used for bounded counts, nor for count processes from the INARMA family. In this paper, we present an alternative solution, named the Tobit approach, for achieving approximate linearity together with negative ACF values, which is more generally applicable than the softplus approach. A Skellam--Tobit INGARCH model for unbounded counts is studied in detail, including stationarity, approximate computation of moments, maximum likelihood and censored least absolute deviations estimation for unknown parameters and corresponding simulations. Extensions of the Tobit approach to other situations are also discussed, including underlying discrete distributions, INAR models, and bounded counts. Three real-data examples are considered to illustrate the usefulness of the new approach.

\end{abstract}

\section{Introduction}
\label{Introduction}
Count time series $x_1,\ldots,x_n$, $n\in\bbn=\{1,2,\ldots\}$, are discrete-valued quantitative time series, which stem from an underlying count process $\{X_t\ |\ t\in\bbz\}$, $\bbz=\{\ldots,-1,0,1,\ldots\}$, with the range being contained in $\bbn_0=\{0,1,\ldots\}$. More precisely, if the range of $\{X_t\}$ is equal to~$\bbn_0$, we are concerned with unbounded counts, whereas bounded counts have the finite range $\{0,\ldots,N\}$ with some upper bound $N\in\bbn$ \citep{weiss18}. Several models for count time series have been developed during the last decades, often inspired by the traditional autoregressive moving average (ARMA) models for real-valued time series. Such proposals include the integer-valued ARMA (INARMA) models for stationary count time series \citep[among others by][]{kenzie85,alzaid90,duli91}, where the ARMA recursion is adapted by using types of ``thinning operators'', and the integer-valued generalized autoregressive conditional heteroscedasticity (INGARCH) models \citep[among others by][]{ferland06,fokianos09,xu12,zhu12}, which are conditional regression models having a linear conditional mean. Both the INARMA and the INGARCH models exhibit an ARMA-like autocorrelation structure, in the sense that their autocorrelation function (ACF) satisfies a set of Yule--Walker equations. However, by contrast to the ordinary ARMA models, the attainable range of ACF~values of these ARMA-like count models is often restricted to only positive ACF values. This follows from the parameter constraints being needed to ensure the non-negative outcomes for the process.

To achieve negative ACF values within the class of conditional regression models, a traditional solution is to use a log~link \citep{fokianos11}, but then the model behaves distinctly non-linear and the ARMA-like ACF is lost.
Linear models, by contrast, are popular among practitioners as they are ``simple, useful, and interpretable in a wide range of contexts'' \citep[p.~479]{grunwald00}. They also allow to describe important stochastic properties by closed formulae (\eg the ACF by Yule--Walker equations) and to derive simple moment estimators for initial model fit.
Thus, \citet{weissetal22} recently proposed the family of softplus INGARCH models, which use the softplus function as a smooth but nearly (piecewise) linear response function. This model successfully combines an approximately ARMA-like ACF with the ability to handle negative ACF values. However, the softplus approach is limited to the INGARCH family for unbounded counts, \ie it can neither be used for bounded counts, nor for count processes from the INARMA family.

Therefore, in this paper, we present an alternative solution for achieving approximate linearity together with negative ACF values, which is more generally applicable than the softplus approach. For comparison purposes, our novel approach is first exemplified and discussed within the INGARCH family for unbounded counts, but later we show how to adapt it to other situations as well. Thus, as the first step, we propose a novel family of conditional regression models for unbounded counts, which is defined as a combination of the INGARCH approach with the so-called Tobit model (\emph{To}bin's pro\emph{bit} model, see \citet{tobin58,goldberger64} as well as Section~\ref{Tobit INGARCH}).
An INGARCH$(p,q)$ model with $p\geq 1$ and $q\geq 0$ is characterized by requiring the conditional mean $M_t=E(X_t|\mathcal{F}_{t-1})$, where $\mathcal{F}_{t-1}$ is the $\sigma$-field generated by $\{X_{t-1}, X_{t-2},\ldots\}$, to be a linear expression in the last $p$ observations and the last $q$ conditional means (``feedback terms''), \ie
\ba
\label{INGARCHmodel}
 M_t\ =\ a_0\ +\ \sum_{i=1}^{p}\, a_i\, X_{t-i}\ +\ \sum_{j=1}^{q}\, b_j\, M_{t-j}.
\ea
In standard INGARCH models, see Appendix~\ref{INGARCH} for details,
$M_t$ is used as the conditional mean of a count random variable and, thus, has to be a positive real number. Therefore, recursion \eqref{INGARCHmodel} is accompanied by the constraints $a_0>0$ and $a_1,\ldots,a_{p},b_1,\ldots,b_{q}\geq 0$, which, in turn, cause the ACF to be non-negative. So the first idea to allow for negative ACF values might be to skip the aforementioned parameter constraints and to use $\max\big\{0, M_t\big\}$ as the conditional mean of the generated count. Since the response function $\max\{0,x\}$ is also known as the rectified linear unit (ReLU) function, let us refer to this approach as the \emph{ReLU INGARCH model}. This model, however, has a serious drawback: if $M_t\leq 0$ happens at time~$t$, then the next count generated by the ReLU INGARCH model has the mean zero, \ie we have a degenerate count distribution with all probability mass in zero. This, in turn, will cause problems in likelihood computations, time series forecasting, and others. Therefore, a different solution is required to allow~$M_t$ according to \eqref{INGARCHmodel} to become negative.

The above drawbacks of the ReLU INGARCH model are avoided by our novel \emph{Tobit INGARCH model}. There, we first generate an integer-valued random variable~$X_t^*$ with range~$\bbz$ (so including negative integers!) and having the conditional mean~$M_t$ (which is possibly negative). $X_t^*$ might be understood as a kind of ``auxiliary'' or ``intermediate'' variable, which is only required to catch the case where the negative dependence parameters in \eqref{INGARCHmodel} cause a negative outcome. Such negative outcomes are undesirable as we aim at modeling a count process. Therefore, in a second step, $X_t^*$ is left-censored at zero by computing $X_t = \max\big\{0, X_t^*\big\}$ (Tobit approach); details are given in Section~\ref{Tobit INGARCH}. Note that our Tobit approach differs from the aforementioned ReLU approach, because the counts generated by $X_t = \max\big\{0, X_t^*\big\}$ always have a truly positive mean, \ie we never have a degenerate count distribution (provided that $X_t^*$ takes positive values with non-zero probability). A detailed discussion of this general model as well as of its stochastic properties is provided by Section~\ref{Tobit INGARCH}. Afterwards in Section~\ref{Skellam--Tobit INGARCH}, we consider the important special case, where~$X_t^*$ is generated from a conditional Skellam distribution \citep[Section~4.12.3]{johnson05}, leading to the Skellam--Tobit INGARCH model. For this model, we also evaluate its abilities for approximating linearity and for capturing negative ACF values. Section~\ref{Estimation} then presents and investigates approaches for parameter estimation.
Section~\ref{Possible Extensions} discusses a variety of possible extensions, \ie how the Tobit approach could be adapted to models outside the unbounded INGARCH class, and  the extensions are mainly about underlying discrete distributions, INAR models, and bounded counts.
Section~\ref{Real Applications} illustrates the application of the Tobit count models to three real-world count time series: lottery winners counts,
chemical process yields, and daily air quality levels.
Finally, Section~\ref{Conclusions} concludes and outlines issues for future research.

\section{The Class of Tobit INGARCH Models}
\label{Tobit INGARCH}
The basic Tobit model, a censored probit regression model, was introduced by \citet{tobin58} and has been extended in several ways, see Chapter~6 in \citet{maddala83} for a comprehensive survey.
Some of these extensions refer to time series data. For example, a dynamic Tobit model for real-valued random variables~$Y_t$ (left-censored at zero) might be of the form
\ba
\label{dynamicTobit}
Y_t\ =\ \max\big\{0,\ Y_t^*\big\}
\quad\text{with}\quad
Y_t^*\ =\ \sum_{i=1}^p \rho_i\,Y_{t-i} + \epsilon_t\ +\fgamma^\top\fz_t,
\ea
see \citet{maddala83,honore93,honore04,dejong11}. If~$\epsilon_t$ is normally distributed, then the conditional distribution of~$Y_t^*$ given $Y_{t-1},\ldots$ and~$\fz_t$ is also normal, \ie $Y_t$ with range $[0;\infty)$ is generated by censoring the $\bbr$-valued~$Y_t^*$ at zero. In what follows, this idea is transfered to the censoring of a $\bbz$-valued random variable~$X_t^*$, leading to the count random variable $X_t = \max\big\{0, X_t^*\big\}$ with range~$\bbn_0$.

\bigskip
To overcome the limitations of the ordinary INGARCH model while (approximately) preserving its additive structure, we propose the novel class of Tobit INGARCH models. For defining this model, let us introduce the $\bbz$-valued random operator~$\mathcal S_{\fnui}(\mu)$, where $\mu\in\bbr$ denotes the mean of~$\mathcal S_{\fnui}(\mu)$, and $\fnu$ comprises some further distributional parameters (\eg such as a dispersion parameter).
The letter ``$\mathcal S$'' might be understood as referring to \emph{s}igned integer outcomes from~$\bbz$. Given the values of $\mu$ and $\fnu$, $\mathcal S_{\fnui}(\mu)$ is a random variable on~$\bbz$. The terminology ``operator'' is used in analogy to the thinning operators in INARMA modeling, to express that $\mathcal S_{\fnui}(\mu)$ acts on~$\mu$ by generating an integer value.
It is assumed that the distribution of $\mathcal{S}_{\fnui}(\mu)$ is such that $P\big(\mathcal{S}_{\fnui}(\mu) > 0\big) >0$ holds for all $\mu$ and $\fnu$.
Later in Section~\ref{Skellam--Tobit INGARCH}, we consider the special case of the operator following the Skellam distribution with mean~$\mu$ and dispersion parameter~$\delta>0$, where this assumption is satisfied.

In analogy to \eqref{dynamicTobit}, the \emph{Tobit INGARCH model} is defined by the equation
\ba\label{TobitINGARCHmodel}
X_t\ =\ \max\big\{0,\ X_t^*\big\}
\quad\text{with}\quad
X_t^*\ =\ \mathcal S_{\fnui}(M_t),
\ea
where we assume the conditional mean $M_t=E(X_t^*|\mathcal{F}_{t-1})$ to satisfy the linear INGARCH equation
\ba\label{LinCondMean}
M_t\ =\ \alpha_0+\sum_{i=1}^p\alpha_i\, X_{t-i}+\sum_{j=1}^q\beta_j\, M_{t-j}
\ea
with $\alpha_0,\alpha_1,\ldots,\alpha_p,\beta_1,\ldots,\beta_q\in\bbr$, recall \eqref{INGARCHmodel}. Note that \eqref{TobitINGARCHmodel} is equivalently stated as $X_t = X_t^*\,\indfkt_{\bbn}(X_t^*)$,
because the integer ReLU function $\max\{0,x\}$ can be rewritten as $x\,\indfkt_{\bbn}(x)$, where the indicator function $\indfkt_A(x)$ equals~1 (0) if $x\in A$ ($x\not\in A$).

According to \citet{grunwald00}, a time series model for $\{X_t\}$ can be classified as being linear (thus satisfying, \eg a set of Yule--Walker equations for the ACF) if its conditional mean $E(X_t|\mathcal{F}_{t-1})$ is linear in past observations. From \eqref{TobitINGARCHmodel} and \eqref{LinCondMean}, it gets clear that the $\bbz$-valued~$X_t^*$ has an exactly linear conditional mean and would, thus, satisfy the linearity requirement of \citet{grunwald00}. However, it might happen that~$X_t^* = \mathcal S_{\fnui}(M_t)$ takes a negative value, which contradicts the count nature of the target process. Therefore, \eqref{TobitINGARCHmodel} turns~$X_t^*$ into the count~$X_t$ that we are actually interested in, at the price that~$X_t$ is no longer linear in~$X_{t-i}$ and~$M_{t-j}$ because of the censoring. Nevertheless, we shall see in Section~\ref{On Moments and Approximate Linearity} below that by an appropriate choice of~$\mathcal S_{\fnui}(\cdot)$, we can achieve that $E(X_t|\mathcal{F}_{t-1})$ behaves at least \emph{nearly} linearly in past observations, such that the corresponding Tobit INGARCH model can be applied to linear count time series data in practice.

For being able to evaluate the extent of linearity, let us first derive formulae for the conditional probabilities and moments of $\{X_t\}$. While~$X_t^*$ is a $\bbz$-valued random variable, the~$X_t$ generated by \eqref{TobitINGARCHmodel} is a count random variable with conditional distribution
$$
P(X_t=x|\mathcal{F}_{t-1})\ =\ \left\{\begin{array}{ll}
P\big(\mathcal S_{\fnui}(M_t)=x\big) & \text{if } x>0,\\[.5ex]
P\big(\mathcal S_{\fnui}(M_t)\leq 0\big) & \text{if } x=0.
\end{array}\right.
$$
As a result, the conditional mean of~$X_t$ given~$\mathcal{F}_{t-1}$ computes as
\ba
\label{TINGARCHcmean}
E(X_t|\mathcal{F}_{t-1})
\ =\ E\big(X_t^*\,\indfkt_{\bbn}(X_t^*)\big|\mathcal{F}_{t-1}\big)
\ =\ \sum_{x=1}^\infty x\,P\big(\mathcal S_{\fnui}(M_t)=x\big).
\ea
Higher-order conditional moments are computed analogously as
\ba
\label{TINGARCHcmom}
E(X_t^r|\mathcal{F}_{t-1}) = E\big((X_t^*)^r\,\indfkt_{\bbn}(X_t^*)\big|\mathcal{F}_{t-1}\big)\quad\text{with}\quad
r\in\bbn.
\ea
We note that these moments are equal to the corresponding partial moments of~$X_t^*$ (left-censored at zero). So having specified the $\bbz$-valued distribution of~$\mathcal S_{\fnui}(\mu)$, these partial moments can be derived. This is illustrated in Section~\ref{Skellam--Tobit INGARCH}, where the special case of a conditional Skellam distribution is discussed in detail.

\section{Skellam--Tobit INGARCH Model}
\label{Skellam--Tobit INGARCH}

\subsection{Skellam Distribution and Partial Moments}
\label{Skellam Distribution and Partial Moments}
Probably the most well-known example of a parametric distribution for a $\bbz$-valued random variable is the \emph{Skellam distribution}, $\skellam(\lambda_1,\lambda_2)$ with $\lambda_1,\lambda_2>0$, see Section~4.12.3 in \citet{johnson05} and the references mentioned therein. It is sometimes also referred to as the ``Poisson-difference distribution'', because $X^*\sim\skellam(\lambda_1,\lambda_2)$ can be generated as the difference $X^*=X_1-X_2$ of the two independent Poisson variables $X_1\sim\poi(\lambda_1)$ and $X_2\sim\poi(\lambda_2)$. Its mean and variance are given by $\mu=\lambda_1-\lambda_2 \in\bbr$ and $\sigma^2=\lambda_1+\lambda_2 > |\mu|\geq 0$, respectively. Its probability mass function (PMF) equals
\ba
\label{skellam}
P(X^*=x)\ =\ {\rm e}^{-\lambda_1-\lambda_2}\,\big(\lambda_1/\lambda_2\big)^{x/2}\,I_x\big(2\,\sqrt{\lambda_1\lambda_2}\big)\qquad\text{for } x\in\bbz,
\ea
where
\ba
\label{bessel}
I_n(z)\:=\ \left(\frac{z}{2}\right)^n\sum_{k=0}^{\infty}\frac{(\frac{z}{2})^{2k}}{k!\cdot (k+n)!}\ \text{for } n\in\bbn_0,\qquad
I_{-n}(z)\:=\ I_n(z),
\ea
denotes the modified Bessel function of the first kind, which is implemented in~R by \texttt{besselI}.
This distribution has been used to analyze some independent data (such as \citet{karlis06,karlis09}) and time series data (such as \cite{koopman17}).

\begin{remark}
The modified Bessel function of the first kind satisfies
\ba
\label{bessel2}
I_{n+1}(z)\ =\ I_{n-1}(z)\ -\ \frac{2n}{z}\cdot I_{n}(z).
\ea
In addition, we have
\begin{align}\label{derbessel}
I_n'(z)=\frac{1}{2}[I_{n-1}(z)+I_{n+1}(z)],~~I_n'(z)=I_{n-1}(z)-\frac{n}{z}I_{n}(z),~~
I_n'(z)=\frac{n}{z}I_{n}(z)+I_{n+1}(z),
\end{align}
see \citet[\S\S~17]{jeffrey08} for these and further properties.

Equations \eqref{bessel} and \eqref{bessel2} immediately imply important relations for the Skellam's PMF \eqref{skellam}, namely
\ba\label{skellam2}
P(X^*=-x)
\ \overset{\eqref{bessel}}{=}\ e^{-\lambda_1-\lambda_2}\,\big(\lambda_1/\lambda_2\big)^{x/2-x}\,I_{x}\big(2\,\sqrt{\lambda_1\lambda_2}\big)
\ =\
\big(\lambda_2/\lambda_1\big)^{x}\,P(X^*=x),
\ea
and
\ba\label{skellam3}
\begin{array}{@{}rl}
P(X^*=x+1)
\ \overset{\eqref{bessel2}}{=}&
e^{-\lambda_1-\lambda_2}\,\big(\lambda_1/\lambda_2\big)^{(x+1)/2}\,\Big(I_{x-1}\big(2\,\sqrt{\lambda_1\lambda_2}\big)\ -\ \frac{x}{\sqrt{\lambda_1\lambda_2}}\,I_{x}\big(2\,\sqrt{\lambda_1\lambda_2}\big)\Big)\\[1ex]
\ =&
\big(\lambda_1/\lambda_2\big)\,P(X^*=x-1)\ -\ \big(x/\lambda_2\big)\,P(X^*=x).
\end{array}
\ea
These results shall be useful for computing partial moments, see the details below.
\end{remark}

The cumulative distribution function (CDF) of $\skellam(\lambda_1,\lambda_2)$ is easily computed in practice, by utilizing Equation (4) in \citet{johnson59} on a connection to the non-central $\chi^2$-distribution. Let $\chi^2(\nu, \tau)$ denote the non-central $\chi^2$-distribution with $\nu$ degrees of freedom and non-centrality parameter~$\tau$ (whose CDF is provided by \texttt{pchisq} in R). Then, Equation (4) in \citet{johnson59} implies that
\ba
\label{skellam_cdf}
P(X^*\leq x)\ =\ \left\{\begin{array}{ll}
P(Q\leq 2\lambda_2) \text{ with } Q\sim\chi^2(-2x, 2\lambda_1) & \text{if } x\leq 0,\\[1ex]
1-P(Q\leq 2\lambda_1) \text{ with } Q\sim\chi^2\big(2(x+1), 2\lambda_2\big) & \text{if } x> 0.
\end{array}\right.
\ea
Finally, let us turn to the computation of partial moments. In view of \eqref{TINGARCHcmean}, for example, the partial moment $E\big(X^*\,\indfkt_{\bbn}(X^*)\big)$ is relevant to us, while the second-order moment $E\big((X^*)^2\,\indfkt_{\bbn}(X^*)\big)$, also see \eqref{TINGARCHcmom}, is used for computing the partial variance.
Note that the Skellam distribution is characterized by the Stein equation
\ba
\label{Skellam_Stein}
E\big(X^*\,f(X^*)\big)\ =\
\lambda_1\,E\big(f(X^*+1)\big)\ -\
\lambda_2\,E\big(f(X^*-1)\big)
\ea
for any bounded function $f:\bbz\to\bbr$, see Theorem~1 in \citet{balachandran17}. Such a Stein equation can be utilized to computing sophisticated moment expressions, see \citet{weissaleksandrov22}, and it is now applied for deriving the relevant partial moments.

\begin{proposition}
\label{propSkellamPartmom}
Let $X^*\sim\skellam(\lambda_1,\lambda_2)$, then it has the partial mean
\ba
\label{Skellam_partial}
E\big(X^*\,\indfkt_{\bbn}(X^*)\big)
\ =\
\mu\,P(X^*\geq 0)\ +\
\lambda_2\,\big(P(X^*=0) + P(X^*=1)\big)
\ea
and the second-order partial moment
\ba
\label{Skellam_pmom2}
E\big((X^*)^2\,\indfkt_{\bbn}(X^*)\big)
\ =\
(\sigma^2+\mu^2)\,P(X^*\geq 1)
\ +\
\lambda_2\,\mu\,P(X^*= 1)
\ +\
\lambda_1\,(1+\mu)\,P(X^*=0).
\ea
\end{proposition}
The proof of Proposition~\ref{propSkellamPartmom} is provided by Appendix~\ref{Proof of Proposition propSkellamPartmom}.
The partial-moment expressions in Proposition~\ref{propSkellamPartmom} are evaluated by using Equations \eqref{skellam} and \eqref{skellam_cdf} on the Skellam's PMF and CDF, respectively.
Note that \eqref{Skellam_partial} and \eqref{Skellam_pmom2} imply the variance of~$X^*\,\indfkt_{\bbn}(X^*)$.

In practice, it may be advantageous to reparametrize $\skellam(\lambda_1,\lambda_2)$ by the mean~$\mu$ and an additional dispersion parameter~$\delta>0$, defined by the additive decomposition $\sigma^2=|\mu|+\delta$.
Thus, the reparametrized version $\skellam^*(\mu,\delta)$ is related to the original version $\skellam(\lambda_1,\lambda_2)$ by
\ba
\label{Skellam_reparam}
\lambda_1\ =\ \tfrac{1}{2}\,\big(|\mu|+\mu+\delta\big),\qquad
\lambda_2\ =\ \tfrac{1}{2}\,\big(|\mu|-\mu+\delta\big).
\ea
Note that the limit $\delta\to 0$ leads to (the negative of) a $\poi\big(|\mu|\big)$-variate if $\mu>0$ ($\mu<0$), \ie the Poisson distribution is a boundary case of the Skellam distribution.

\begin{example}
\label{example_mu0}
In the case $\mu=0$ (\ie $\lambda_1=\lambda_2=\delta/2$), the Skellam distribution is symmetric around~0 with $P(X^*=x) = e^{-\delta}\,I_x(\delta)$ and $P(X^*\geq 1) = \big(1-P(X^*=0)\big)/2$. Equations \eqref{Skellam_partial} and \eqref{Skellam_pmom2} simplify to
$$
\begin{array}{rl}
E\big(X^*\,\indfkt_{\bbn}(X^*)\big)\ =&
\tfrac{1}{2}\,\delta\,\big(P(X^*=0) + P(X^*=1)\big)\ =\
\tfrac{1}{2}\,\delta\,e^{-\delta}\,\big(I_0(\delta) + I_1(\delta)\big).
\\[1ex]
E\big((X^*)^2\,\indfkt_{\bbn}(X^*)\big)
\ =&
\delta\,P(X^*\geq 1)
\ +\
\tfrac{1}{2}\,\delta\,P(X^*=0)
\ =\
\tfrac{1}{2}\,\delta.
\end{array}
$$
Thus, we have
$$
\begin{array}{rl}
V\big(X^*\,\indfkt_{\bbn}(X^*)\big)\ =&
\tfrac{1}{2}\,\delta - \tfrac{1}{4}\,\delta^2\,e^{-2\,\delta}\,\big(I_0(\delta) + I_1(\delta)\big)^2.
\end{array}
$$
\end{example}

\subsection{Skellam--Tobit INGARCH Model: Definition and Properties}
\label{Definition and Properties}
Let us define the Skellam operator $\mathcal S_{\fnui}(\mu)$ to generate a random integer number according to $\skellam(\lambda_1,\lambda_2)$, where $\fnu=(\lambda_1,\lambda_2),\mu=\lambda_1-\lambda_2$. Then, adapting the general Tobit INGARCH approach \eqref{TobitINGARCHmodel}, we define the \emph{Skellam--Tobit INGARCH (STINGARCH) model} by
\ba\label{STINGARCHmodel}
X_t\ =\ \max\big\{0,\ X_t^*\big\}
\quad\text{with}\quad
X_t^*\ =\ \mathcal S_{\fnui}(M_t),
\ea
where $M_t=E(X_t^*|\mathcal{F}_{t-1})$ satisfies the linear INGARCH equation \eqref{LinCondMean}. Note that \cite{doukhan21} considered a Skellam GARCH model for $\mathbb{Z}$-valued time series based on modelling the conditional variance,  but here we will provide a method for analyzing non-negative count time series based on modeling the conditional mean.

\begin{theorem}
\label{thmExistence}
Consider the STINGARCH process $\{X_t\}$ defined in \eqref{STINGARCHmodel}. If
\ba\label{st-cond}
\sum_{i=1}^{p}\max\{0,\alpha_i\}+ \sum_{j=1}^{q}|\beta_j|<1,
\ea
then $\{X_t\}$ exists and is stationary.
\end{theorem}

The proof of Theorem~\ref{thmExistence} is provided by Appendix~\ref{Proof of Theorem thmExistence}.

\begin{remark}
The above condition is comparable to the ones in other real-valued dynamic Tobit models, such as \cite{dejong11}, \cite{michel18}, and \cite{bykhovskaya23}. This condition is also comparable to the one for the INGARCH model in \cite{weissetal22}.
Due to the censoring  mechanism, we do not need to impose lower bounds on the parameters $\alpha_i$, $i=1,\ldots,p$.
\end{remark}

In what follows, we consider the STINGARCH model based on the additive reparametrization \eqref{Skellam_reparam}, \ie
\ba\label{STINGARCHmodel_add}
X_t\ =\ \max\big\{0,\ X_t^*\big\}
\quad\text{with}\quad
X_t^*\ =\ \mathcal S_{\delta}(M_t).
\ea
Then, the conditional probability of $X_t^\ast$ given $M_t$ is
\begin{align}\label{conprob}
P(X_t^\ast=x|M_t)&={\rm e}^{-|M_t|-\delta}\left(\frac{|M_t|+M_t+\delta}{|M_t|-M_t+\delta}\right)^{x/2}I_x\Big(\sqrt{\big(|M_t|+M_t+\delta\big)\big(|M_t|-M_t+\delta\big)}\Big)\nonumber\\
&=\left\{\begin{array}{ll}
\displaystyle{\rm e}^{-M_t-\delta} \left(\frac{2M_t+\delta}{\delta}\right)^{x/2}\, I_x\Big(\sqrt{\delta\,(2M_t+\delta)}\Big) & \text{if } M_t\geq0,\\
\displaystyle{\rm e}^{M_t-\delta} \left(\frac{\delta}{-2M_t+\delta}\right)^{x/2}\, I_x\Big(\sqrt{\delta\,(-2M_t+\delta)}\Big) & \text{if } M_t<0.
\end{array}\right.
\end{align}
For the reparametrized STINGARCH model \eqref{STINGARCHmodel_add},
the limit $\delta\to 0$ leads to the ReLU INGARCH model with conditional Poisson distribution $\poi\big(\max\{0,M_t\}\big)$.

\begin{remark}
As explained before, we use an additive reparametrization for the Skellam's relation between variance and mean. This differs from \citet{carallo20}, who use the multiplicative decomposition $\sigma^2=\tilde{\delta}\,|\mu|$ with $\tilde{\delta}>1$ (as $|\mu|<\sigma^2$ always holds). However, to guarantee $\sigma^2>0$, such a multiplicative parametrization requires to assume that $\lambda_1\neq\lambda_2$, which is unnecessarily restrictive. It would also lead to rather different moment plots than those in Figure~\ref{figskellam} in Section~\ref{On Moments and Approximate Linearity} below.

It should be noted that the considered reparametrization does not affect Theorem~\ref{thmExistence}, because we established the existence and stationarity for the original version of Skellam.
\end{remark}

\begin{figure}[t]
\small
\includegraphics[viewport=0 15 335 305, clip=, scale=0.45]{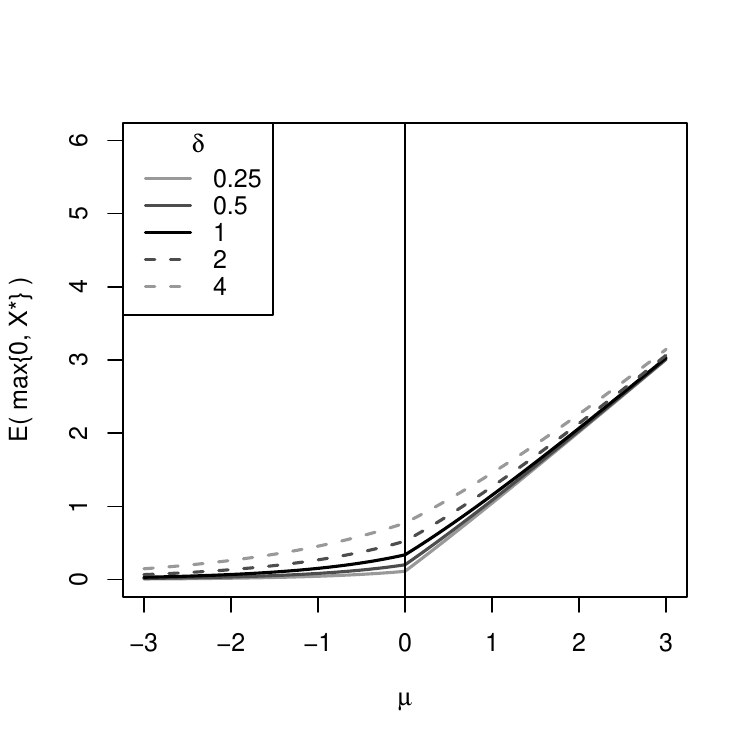}
\includegraphics[viewport=0 15 335 305, clip=, scale=0.45]{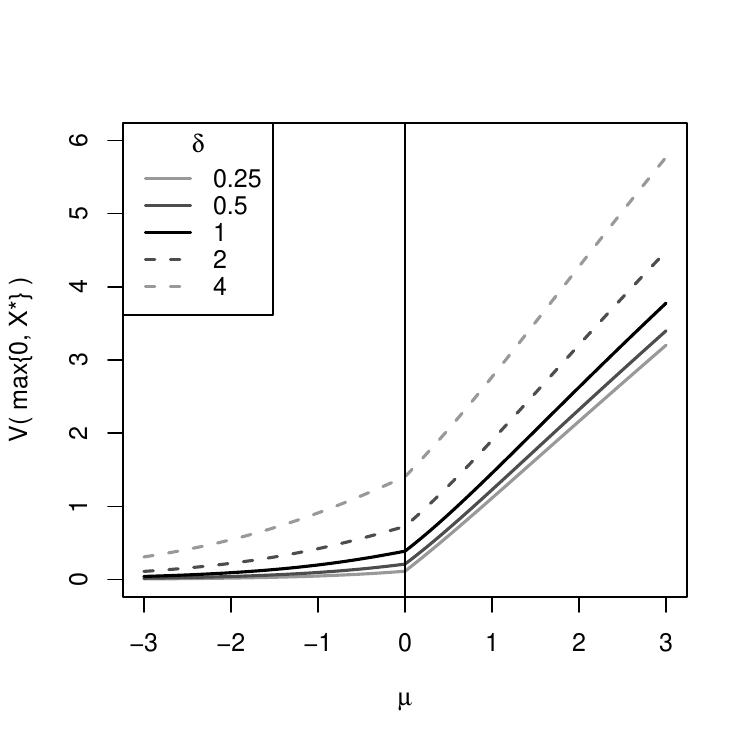}
\includegraphics[viewport=0 15 335 305, clip=, scale=0.45]{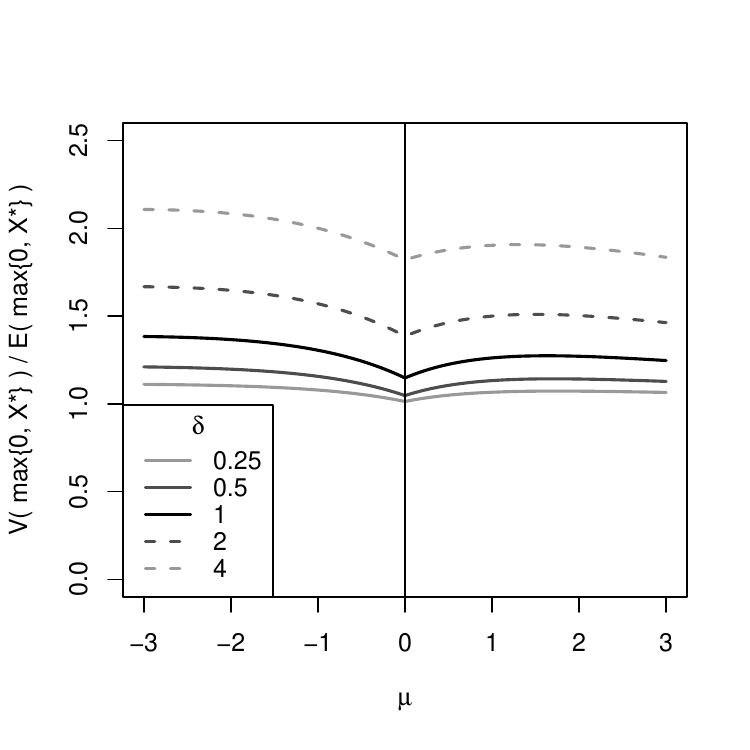}
\caption{Plots (against~$\mu$) of mean, variance, and dispersion ratio (from left to right) of left-censored $\skellam^*(\mu,\delta)$-variate.}
\label{figskellam}
\end{figure}

\subsection{On Moments and Approximate Linearity}
\label{On Moments and Approximate Linearity}
The Tobit INGARCH approach was proposed as an approximate solution for achieving a conditionally linear count model allowing for negative ACF values. This conjecture shall now to be verified.
To compute the conditional moments \eqref{TINGARCHcmean} and \eqref{TINGARCHcmom} for the STINGARCH model \eqref{STINGARCHmodel_add}, the Skellam's partial moments in Proposition~\ref{propSkellamPartmom} are required.
The resulting mean, variance, and dispersion ratio of such left-censored $\skellam^*(\mu,\delta)$-variates are illustrated by Figure~\ref{figskellam}. It can be seen that both the mean and the variance behave roughly linear outside $\mu=0$, and the extent of linearity increases with decreasing~$\delta$. While the mean is rather stable for varying values of~$\delta$, the dispersion ratio increases with increasing~$\delta$ (but is roughly constant for varying~$\mu$). For $\delta\to 0$, the dispersion ratio converges to the constant~1 (for all~$\mu$), whereas the limiting mean (=limiting variance) equals $\max\{0,\mu\}$. This goes along with our above statement that the limit $\delta\to 0$ leads to the Poisson ReLU INGARCH model.

The approximate (piecewise) linearity of the left-censored Skellam's mean in Figure~\ref{figskellam} (for small~$\delta$ such as 0.25) leads us to conjecture that also the STINGARCH model behaves similar to a truly linear model, because it implies that the conditional mean $E(X_t|\mathcal{F}_{t-1})$ is approximately linear in past observations, also recall \citet{grunwald00}. More precisely, we conjecture that the marginal mean~$\mu$, variance~$\sigma^2$, and ACF $\rho(h)$ with time lag $h\in\bbn$ of a STINGARCH process $\{X_t\}$ can be computed in good approximation according to the same formula as for the truly linear, so-called ``dispersed INGARCH'' (DINGARCH) model of \citet{xu12}, see Appendix~\ref{INGARCH} for details, but by also allowing for negative parameter values (which are impossible for the DINGARCH model). Here, the DINGARCH's dispersion parameter~$\eta$ is chosen to equal the dispersion ratio $I_{\mu,\delta}^* := V\big(X^*\,\indfkt_{\bbn}(X^*)\big)\,\big/\,E\big(X^*\,\indfkt_{\bbn}(X^*)\big)$ for $X^*\sim\skellam^*(\mu,\delta)$, which is computed according to Proposition~\ref{propSkellamPartmom}.
Let us illustrate this approximate approach for the STINARCH$(1)$ and STINGARCH$(1,1)$ model. We propose the approximate scheme
\ba
\label{approxMom}
\begin{array}{l@{\qquad}l}
(p,q)=(1,0): &\displaystyle
\mu\ \approx\ \frac{\alpha_0}{1- \alpha_1},
\quad
\frac{\sigma^2}{\mu}\ \approx\ I_{\mu,\delta}^*\cdot\frac{1}{1-\alpha_1^2},
\quad
\rho(h)\ \approx\ \alpha_1^h \text{ for } h\geq 1,
\\[3ex]
(p,q)=(1,1): &\displaystyle
\mu\ \approx\ \frac{\alpha_0}{1- \alpha_1-\beta_1},
\quad
\frac{\sigma^2}{\mu}\ \approx\ I_{\mu,\delta}^*\cdot\frac{1-(\alpha_1+\beta_1)^2+\alpha_1^2}{1-(\alpha_1+\beta_1)^2},
\\[2ex] &\displaystyle
\rho(h)\ \approx\ (\alpha_1+\beta_1)^{h-1}\, \frac{\alpha_1\, \big(1-\beta_1(\alpha_1+\beta_1)\big)}{1-(\alpha_1+\beta_1)^2+\alpha_1^2} \text{ for } h\geq 1,
\end{array}
\ea
for moment computation in these cases. To check the quality of this approximation and, thus, the extent of linearity of the STINGARCH models, we compare the linear moment values being computed according to \eqref{approxMom} with their true values, which are computed numerically exactly for the STINARCH$(1)$ model (by utilizing its Markov property), and approximately for STINGARCH$(1,1)$ (based on a simulated sample path of length~$10^6$).
Results are tabulated in Appendix~\ref{Tables}; let us start with the STINARCH$(1)$ case. For different choices of $\alpha_0,\alpha_1,\delta$, the true values (columns ``tob'') and anticipated linear values (columns ``lin'', computed via \eqref{approxMom}) for marginal mean, dispersion ratio, and partial ACF (PACF) at lags $h=1,2,3$ have been computed, where $\delta=0$ corresponds to the limiting case of the Poi-ReLU INARCH$(1)$ model. The latter just serves as the benchmark for the maximal extent of linearity within the INARCH$(1)$ approach, but it is not relevant for practice because of the drawbacks outlined in Section~\ref{Introduction}.

Table~\ref{tabMom5} provides results for the case, where the marginal mean is close to~5 (``low counts''). Comparing the ``tob-columns'' to the ``lin-columns'', we generally see a good agreement, which further improves for decreasing~$\delta$ (and which is even better for larger mean, see Table~\ref{tabMom10} for comparison). Except for strong negative autocorrelation (as implied by $\alpha_1=-0.75$), the choice $\delta=0.25$, for example, leads to linearity in close approximation. For $\alpha_1=-0.75$, it should be noted that even the exactly piecewise-linear Poi-ReLU INARCH$(1)$ model causes departures from linearity, so some extent of non-linearity cannot be avoided in this extreme situation. It is also worth pointing out that the tob- and lin-values of the dispersion ratio $\sigma^2/\mu$ agree quite well even for larger~$\delta$ (except the extreme scenario $\alpha_1=-0.75$), \ie the approximation $I_{\mu,\delta}^*/(1-\alpha_1^2)$ generally shows a good performance.

If we now turn to the corresponding STINGARCH$(1,1)$ results in Tables~\ref{tabMom115}--\ref{tabMom1110}, where the ACF instead of the PACF is considered, we again see a very close agreement throughout. The additional feedback term did not affect the closeness to linearity compared to our previous STINARCH$(1)$ conclusions.
Thus, to sum up, if~$\delta$ is chosen sufficiently small such as $\delta=0.25$, the novel Tobit approach leads to approximately linear count time series models that capture a wide range of negative ACF values. Therefore, STINGARCH models are recommended as a kind of ``workhorse'' for linear count processes with signed ACF values.

\section{Estimation}
\label{Estimation}
Let $X_1,\ldots,X_n$ be observations from model \eqref{STINGARCHmodel}.
Define the parameter vector $\ftheta=(\alpha_0,\alpha_1,\ldots$, $\alpha_p,\beta_1,\ldots,\beta_q,\delta)^\top$, and let $\fvtheta=(\alpha_0,\alpha_1,\ldots,\alpha_p,\beta_1,\ldots,\beta_q)^\top$, \ie $\ftheta=(\fvtheta^\top,\delta)^\top$.
Let $\fvtheta_0=(\alpha_{00},\alpha_{10},\ldots,\alpha_{p0},\beta_{10},\ldots,\beta_{q0})^\top$ be the true value of $\fvtheta$, and $\ftheta_0=(\fvtheta_0^\top,\delta_0)^\top$ be the one of~$\ftheta$. The parameter space of $\ftheta$ is $\fTheta$ and the parameter space of $\fvtheta$ is $\fTheta^\ast$, which are compact subsets of $\mathbb{R}^{p+q+2}$ and $\mathbb{R}^{p+q+1}$, respectively, and can be further refined when deriving large-sample properties of the following considered estimators.
We have to distinguish two general estimation scenarios:

\begin{description}
\item[Scenario 1:] $\delta$ is considered as a tuning parameter that is specified in advance to achieve approximate linearity. Recalling the discussion in Section~\ref{On Moments and Approximate Linearity}, setting $\delta:=0.25$ or smaller might be such a choice. Then, parameter estimation refers to solely $\fvtheta$.

\item[Scenario 2:] $\delta$ is considered as a parameter of the data-generating process (DGP). Then, parameter estimation refers to whole $\ftheta$.
\end{description}
In our opinion, scenario~1 is more relevant for practice. In view of George Box' famous words ``all models are wrong, but some are useful'', we consider the STINGARCH model as tool (workhorse) for describing count time series exhibiting linear sample properties and negative dependencies. Here, the parameter~$\delta$ is used as a tuning parameter set by the user to control the extent of linearity, similar to the parameter~$c$ of the softplus INGARCH model of \citet{weissetal22} mentioned in Section~\ref{Introduction}. The user may also define multiple candidate models with different extents of linearity. But for completeness, we also investigate scenario~2 in some detail.

We consider two estimation methods: the maximum likelihood estimator (MLE) in Section~\ref{Maximum likelihood estimation}, and the censored least absolute deviations estimator (CLADE) in Section~\ref{Censored least absolute deviations estimation}.
The second method is expected to be robust if the underlying distribution is mis-specified, and it can be applied to estimation scenario~1. The first method can be used for both estimation scenarios.

\subsection{Maximum Likelihood Estimation}
\label{Maximum likelihood estimation}
First, we consider the MLE of $\ftheta_0$ (\ie estimation scenario~2).
Based on \eqref{conprob}, define $p_t(x)=P(X_t^\ast=x|M_t)$ and $f_t(x)=P(X_t^\ast\leq x|M_t)$. Then, the conditional log-likelihood function is
\ba
\label{MLE}
L(\ftheta)=\sum_{t=1}^n l_t(\ftheta)
\quad\text{with}\quad
l_t(\ftheta)\ =\ \indfkt_{\bbn}(X_t)\,\log{p_t(X_t)}\ +\ \indfkt_{\{0\}}(X_t)\,\log{f_t(0)},
\ea
where Equations \eqref{skellam} and \eqref{skellam_cdf} are used for the PMF and CDF computations, respectively.
If $X_t>0$, then
\begin{align*}
l_t(\ftheta)\ =\
\left\{\begin{array}{ll}
\displaystyle-M_t-\delta +\tfrac{1}{2}\,X_t\,\Big[\log(2M_t+\delta)-\log\delta\Big]+\log I_{X_t}\Big(\sqrt{\delta(2M_t+\delta)}\Big) & \text{if } M_t\geq0,
\\[1ex]
\displaystyle M_t-\delta-\tfrac{1}{2}\,X_t\,\Big[\log(-2M_t+\delta)-\log\delta\Big]+\log I_{X_t}\Big(\sqrt{\delta(-2M_t+\delta)}\Big) & \text{if } M_t<0.
\end{array}\right.
\end{align*}
If $X_t=0$, then $l_t(\ftheta)=\log{f_t(0)}$.
The MLE $\hat\ftheta$ is defined as $\hat\ftheta={\arg\max}_{\fthetai} L(\ftheta)$.
If MLE is applied to estimation scenario~1, then~$\delta$ is specified in advance, and the maximization of \eqref{MLE} is done with respect to~$\fvtheta$.

\smallskip
The above likelihood function is cumbersome (for both scenarios), so we need some numerical algorithms to find the MLE $\hat\ftheta$.
The score function and the Hessian matrix are defined as
$$
\fS_n(\ftheta)\ =\ \sum_{t=1}^n \frac{\partial l_t(\ftheta)}{\partial\ftheta}
\qquad\text{and}\qquad
\mH_n(\ftheta)\ =\ -\sum_{t=1}^n \frac{\partial^2 l_t(\ftheta)}{\partial\ftheta\partial\ftheta^\top},
$$
respectively. The MLE $\hat{\ftheta}$ also is a solution to the equation $\fS_n(\ftheta)=\0$.
Detailed expressions for the partial derivatives are discussed in Appendix~\ref{Derivatives of Log-Likelihood Function}.

\smallskip
For establishing consistency and asymptotic normality of MLE, one could verify the regularity conditions in Theorems 2.7 and 3.1 in \cite{newey94} or Theorems 4.1.1 and 4.1.3 in \cite{amemiya85}. But because of the complex expressions for the modified Bessel function of the first kind in the likelihood function, it is hardly possible to do so. Note that the aforementioned theorems would imply that the asymptotic variance of the MLE is $\mU^{-1}\mV\mU^{-1}/n$, where
$$
\mV\ =\ E\left(\frac{\partial l_t(\ftheta)}{\partial\ftheta}\,\frac{\partial l_t(\ftheta)}{\partial\ftheta^\top}\right),\qquad
\mU\ =\ -E\left(\frac{\partial^2 l_t(\ftheta)}{\partial\ftheta\partial\ftheta^\top}\right),$$
and where these two matrices are evaluated at the true value $\ftheta=\ftheta_0$. This implies to approximate the asymptotic variance by $(\hat{\mU}_n\hat{\mV}^{-1}_n\hat{\mU}_n)^{-1}/n$,
where
$$
\hat{\mV}_n\ =\ \left.\frac{1}{n}\sum_{t=1}^n\frac{\partial l_t(\ftheta)}{\partial\ftheta}\,\frac{\partial l_t(\ftheta)}{\partial\ftheta^\top}\right|_{\fthetai=\hat\fthetai},\qquad
\hat{\mU}_n\ =\ -\left.\frac{1}{n}\sum_{t=1}^n\frac{\partial^2 l_t(\ftheta)}{\partial\ftheta\partial\ftheta^\top}\right|_{\fthetai=\hat\fthetai}.
$$
As a practically feasible solution, we recommend to approximate the MLE's standard errors by using the numerical Hessian of the maximized log-likelihood function. More precisely, we use the square-roots of the diagonal of the inverse Hessian, as it is common practice in applications of the ML~approach. The performance of these approximate standard errors is later investigated by simulations in Section~\ref{Results from Simulation Study}.

\subsection{Censored Least Absolute Deviations Estimation}
\label{Censored least absolute deviations estimation}
CLADE is a popular estimation method for censored models since \citet{powell84}, also see \citet{dejong11}, \citet{bilias19} and \cite{bykhovskaya23}.
As before, let $\fvtheta=(\alpha_0,\alpha_1,\ldots$, $\alpha_p,\beta_1,\ldots,\beta_q)^\top$, and let $\fvtheta_0=(\alpha_{00},\alpha_{10},\ldots,\alpha_{p0},\beta_{10},\ldots,\beta_{q0})^\top$ be its true value.
The CLADE $\check\fvtheta$ is formalized as
\begin{equation}\label{clade}
\check\fvtheta=\arg\min_{\fvthetai\in\fThetai^\ast}\sum_{t=1}^n \big|X_t-\max\{0, M_t\}\big|,
\end{equation}
where $M_t$ is defined in \eqref{LinCondMean}. Here, we consider the CLADE for $\fvtheta_0$, not for $\ftheta_0$, because this estimator does not provide a first-round estimate for $\delta$.

For real-valued dynamic Tobit models, \cite{dejong11} established consistency and asymptotic normality of CLADE based on \citet{powell84}, and \cite{bykhovskaya23} considered CLADE in a more complex case. But their analyses cannot be generalized here, because their techniques use a error density, which is absent for our model.
This problem deserves further study in future research.

\begin{remark}
In general, the minimization of absolute deviations in models with a positive part is a non-convex problem, and
designing numerical algorithms requires special care.
The method in \citet{bilias19} allows to find an exact computation of $\check\fvtheta$, \ie the CLADE minimization problem \eqref{clade} can be reformulated as the following mixed integer programming problem:
\begin{align*}
  &\min_{\{\fvthetai,(\gamma_t,\phi_t,sm_t,sp_t,t=1,\ldots,n)\}}\sum_{t=1}^n(sm_t+sp_t)\quad\text{subject to}\\
  &X_t-\phi_t+sm_t-sp_t=0\\
	&\phi_t\geq \max\{0,\ M_t\},\\
  &\phi_t\leq \min\big\{M_t+\omega_t(1-\gamma_t),\ \omega_t\gamma_t\big\},\\
  &sm_t\geq0,\quad sp_t\geq0,\quad\gamma_t\in\{0,1\},
\end{align*}
where $t=1,\ldots,n$. Here, $\omega_t$ is an arbitrary parameter that must be sufficiently large such that a solution to the system of
inequalities exists. In other words, the inequalities only represent the max-operator exactly if $\omega_t$ is large enough, specifically
if $\omega_t>|M_t|$.
The auxiliary choice variables $sm_t$ and $sp_t$ are the under-achievement and over-achievement positive slack variables for the
CLAD model, and the positive variable $\phi_t$ is equal to $\max\{0, M_t\}$. The optimization
is done with respect to the binary choice variables $\gamma_t$ and the continuous choice variables $sm_t$, $sp_t$, $\phi_t$, and $\fvtheta$.
See \citet{bilias19} for more details about the method.
\end{remark}

For the STINARCH case (\ie all $\beta_j$s in \eqref{LinCondMean} are zero, so $M_t\ =\ \alpha_0+\sum_{i=1}^p\alpha_i\, X_{t-i}$ and $\fvtheta=(\alpha_0,\alpha_1,\ldots$, $\alpha_p)^\top$),
a possible alternative to \eqref{clade} could be to use squared deviations instead of absolute deviations, thus leading to a conditional least squares (CLS) estimator.
The  ordinary CLS estimator, however, obtained by minimizing $\sum_{t=1}^n \big(X_t-M_t\big)^2$ over $\fvtheta\in\fTheta^\ast$,
will not lead a consistent estimator, which can be seen from a simple example: If $p=q=0$ with $M_t=\alpha_0$, then we have the estimator
$\tilde\alpha_0=\sum_{t=1}^nX_t/n$. Hence, $\tilde\alpha_0$ converges to $E(X_t)\geq0$ using the law of large numbers, but $\alpha_0\in\mathbb{R}$, so $\tilde\alpha_0$ is inconsistent.
A similar discussion for other models can also be found in \cite{bykhovskaya23}.
Therefore, we consider the following CLS estimator~$\breve{\fvtheta}$:
\begin{align*}
\breve{\fvtheta}=\arg\min_{\fvthetai\in\fThetai^\ast}\sum_{t=1}^n \big(X_t-\max\{0, M_t\}\big)^2.
\end{align*}
According to Proposition \ref{propSkellamPartmom}, we know that
$E(X_t|\mathcal{F}_{t-1})\neq M_t$ or $\max\{0,M_t\}$, so the above CLS estimator is not the common one (where the conditional expectation is used to construct the objective function).
Note that $\max\{0,M_t\}\neq M_t\indfkt_{\{X_t>0\}}$, and the objective function is discontinuous with respect to the parameters. So we cannot obtain an explicit expression for the CLS estimator, the theoretic properties of which need to be further studied.
From the simulation experiment in Table~\ref{tabEst1} below, we find that the CLS estimator~$\breve{\fvtheta}$ shows a satisfactory finite-sample performance in practice.

\subsection{Results from Simulation Study}
\label{Results from Simulation Study}
To analyze the performance of the different estimation approaches, we did a comprehensive simulation study. For various parametrizations of STINARCH$(1)$ and STINGARCH$(1,1)$ processes, for different sample sizes~$n$, and for both estimation scenarios, 1,000 stationary count time series were generated. The estimates were determined via numerical optimization (by using R's \texttt{optim}), where for the STINGARCH$(1,1)$ model, we used the initialization $M_0:=\alpha_0$. Relevant results are tabulated in Appendix~\ref{Tables}.
Table~\ref{tabEst1} for the STINARCH$(1)$ process refers to estimation scenario~1, \ie $\delta$ is considered as a tuning parameter and specified in advance as~$0.25$ such that the model behaves nearly linear. In estimation scenario~1, the model is taken as a working tool to handle linear counts, \ie one does not necessarily expect that the DGP is also STINARCH$(1)$ with exactly $\delta=0.25$. Therefore, we considered STINARCH$(1)$ DGPs with true $\delta\in\{0,0.25,1\}$, to check the robustness of estimates with respect to a possible mis-specification of~$\delta$. Table~\ref{tabEst1} shows that in any case, bias and standard errors (SEs) decrease with increasing~$n$, confirming the consistency of the estimates in all cases. But there are also notable differences, especially for low~$n$. For negative~$\alpha_1$, the CLADE~$\check{\alpha}_0$ shows a clear negative bias, whereas the MLE~$\hat{\alpha}_0$ and CLS~$\breve{\alpha}_0$ exhibit an only mild bias. For positive~$\alpha_1$, the situation is reversed, the MLE~$\hat{\alpha}_0$ and CLS~$\breve{\alpha}_0$ have a notable positive bias. The $\alpha_1$-estimates, by contrast, have quite similar properties for all estimation approaches and parametrizations.
Table~\ref{tabEst1} is complemented by Table~\ref{tabEst1b}, where the performance of the MLE's approximative SE (derived from the log-likelihood's Hessian) is investigated. Especially for positive~$\alpha_1$, we see a rather good agreement between the simulated SEs and the mean of the approximated SEs. For negative~$\alpha_1$ and small~$n$, we observe a slight exceedance of the true SE, but the discrepancy vanishes with increasing~$n$. It is also worth noting that the computation of the approximate SEs was always possible in the case of positive~$\alpha_1$, while the Hessian was not invertible for negative~$\alpha_1$ in 1--2\,\% of all simulation runs.

Table~\ref{tabEst1delta} again considers the STINARCH$(1)$ model, but now regarding estimation scenario~2, \ie $\delta$ is estimated together with $(\alpha_0,\alpha_1)$. In this case, only MLE is applicable. It can be seen that the bias and SE properties of~$(\hat{\alpha}_0,\hat{\alpha}_1)$ are essentially the same as if~$\delta$ is specified (recall Table~\ref{tabEst1}), but the performance of~$\hat{\delta}$ is interesting. Although~$\hat{\delta}$ is visibly biased for, say, $n\leq 250$, the bias decreases and becomes negligible for $n\geq 500$. Also the standard errors of~$\hat{\delta}$ decrease with increasing~$n$, but they are still rather larger even for $n=1000$ (compared to the true value of~$\delta$). These difficulties in estimating~$\delta$ are plausible in view of Section~\ref{On Moments and Approximate Linearity}, where we recognized the nearly linear properties of STINGARCH models for a range of $\delta$-values (impeding the distinguishability of different~$\delta$).
As before, we also analyzed the performance of the approximate SEs. While the approximate SEs still agree reasonably well with the simulated SEs in the mean (also the ones for~$\delta$), the inclusion of~$\delta$ into parameter estimation causes another problem: the percentage of simulation runs where the Hessian cannot be inverted increases considerably in some cases, namely with decreasing~$n$ and~$\delta$. For $\delta=0.25$, it takes values about 40\,\% if $n=100$, but decreases below 10\,\% for $n=1000$. For $\delta=1$, the percentages of non-invertibility are much smaller, namely around 12\,\% for $n=100$ and between 0--2\,\% for $n=1000$.

Let us return to estimation scenario~1 (using~$\delta$ as a tuning parameter), which is more relevant for practice in our opinion. In Tables~\ref{tabEst11}--Table~\ref{tabEst112}, estimation results for STIN\-GARCH$(1,1)$ models are shown. Since MLE and CLS again showed analogous bias and SE properties, we omit the CLS results this time to save some space. Furthermore, we now focus on sample sizes $n\geq 250$, because it is well known for INGARCH-type models that the inclusion of the feedback term~$M_{t-1}$ increases the sample-size requirement for valid estimation. For $(\alpha_1,\beta_1)=(-0.25,0.45)$, the CLADE of~$\alpha_0$ succeeds in terms of bias and SE, but otherwise, there is hardly any clear advantage but often a deterioration compared to MLE. Therefore, taking all results together, the MLE appears to be the preferable estimator for STINGARCH processes. In addition, it allows to approximate the SEs from the log-likelihood's Hessian. As can be seen from Table~\ref{tabEst113}, the simulated SE and the mean of the approximate ones agree quite well in nearly any case (sometimes slight deviations if $n=250$). It is also worth noting that we hardly experienced any problems in inverting the Hessian; if there was a non-zero percentage at all, then it was usually much below 1\,\%.

\section{Possible Extensions}
\label{Possible Extensions}
First, we state some possible extensions based on the unbounded INGARCH modeling framework developed in Section~\ref{Tobit INGARCH}. In Section~\ref{Skellam--Tobit INGARCH}, the Skellam distribution is used as the underlying conditional distribution for illustration, but other choices are possible as well. For examples, if stronger dispersion is required, one my use the generalized Poisson difference distribution discussed in \citet{carallo20} instead, which includes the Skellam distribution as a special case. In addition, unlike \citet{agosto16} where a non-negative link function is used for possible covariates, such covariates could be easily and naturally included in \eqref{LinCondMean}.

\smallskip
Second, the Tobit approach proposed in Section~\ref{Tobit INGARCH} is not only relevant for unbounded INGARCH models, it might also be used for different integer-valued time series models such as those from the INAR (integer-valued autoregressive) family. While the ordinary INAR models for count time series \citep[see][Chapters~2--3]{weiss18} are well-defined only for positive parameter values and can, thus, only describe positive ACF values, also extensions to $\bbz$-valued INAR-type models have been proposed in the literature. The first contribution appears to be the INARS$(p)$ model (S=signed) by \citet{kim08}, which relies on a model recursion of the form
\ba
\label{INARSp}
X_t\ =\ \alpha_1\odot X_{t-1}\ + \ldots +\ \alpha_{p}\odot X_{t-{p}}\ +\ \epsilon_t,
\ea
where ``$\odot$'' denotes the signed binomial thinning operator, and the $(\epsilon_t)$ are \iid\ $\bbz$-valued innovations. ``$\odot$'' is defined by requiring the conditional distribution $\alpha\odot X\big|X$ to be $\textup{sgn}(\alpha)\cdot \textup{sgn}(X)\cdot \bin\big(|X|, |\alpha|\big)$, where $-1<\alpha<1$, where~$X$ is a $\bbz$-valued random variable, and where $\textup{sgn}(z)=1$ for $z\geq 0$ and~$-1$ otherwise. The special case of the INARS$(1)$ model according to \eqref{INARSp} has been investigated by \citet{andersson14} together with Skellam-distributed innovations, whereas \citet{alzaid14} discuss a modification allowing for a Skellam marginal distribution.

Applying the Tobit approach of Section~\ref{Tobit INGARCH} to the INARS$(p)$ model \eqref{INARSp}, we propose the $p$th-order \emph{Tobit INAR model} defined by
\ba
\label{TobitINARp}
X_t\ =\ \max\big\{0,\ X_t^*\big\},\qquad
X_t^*\ =\ \alpha_1\odot X_{t-1}\ + \ldots +\ \alpha_{p}\odot X_{t-{p}}\ +\ \epsilon_t.
\ea
Unlike the ordinary INAR$(p)$ model for count time series \citep[see][]{alzaid90,duli91}, the AR-parameters $\alpha_1, \ldots, \alpha_{p}\in (-1,1)$ are allowed to take negative values (obviously, if all $\alpha_1, \ldots, \alpha_{p}>0$, then \eqref{TobitINARp} reduces to the ordinary INAR$(p)$ recursion).
The observations $\{X_t\}$ in \eqref{TobitINARp} are counts, and~$X_t^*$ is allowed to take negative values only to capture the effect of negative AR-parameters. But as we are actually interested in count time series modeling, we assume the \iid\ innovations $(\epsilon_t)$ to be counts as well.
A more detailed investigation of such Tobit INAR$(p)$ models and their abilities to handle negative autocorrelations appears to be an interesting direction for future research.

\smallskip
Third, another possible extension refers to the case of bounded counts, \ie where the range of $\{X_t\}$ is bounded from above by some $N\in\bbn$. Both INAR- and INGARCH-type models have been proposed for such a case, see \citet{kenzie85,ristic16} as well as the survey in \citet[Chapters~3--4]{weiss18}. Here, the additional difficulty arises from the fact that the range $\{0,\ldots,N\}$ is bounded from two sides. Thus, instead of using the ReLU function for constructing a Tobit model, it suggests itself to use a type of clipped ReLU (cReLU) function \citep[see][]{cai17} instead. Let us focus on the BINGARCH case (B=bounded) for illustration.
Using again the recursive scheme \eqref{LinCondMean} for the conditional mean $M_t=E(X_t^*|\mathcal{F}_{t-1})$, the \emph{Tobit BINGARCH model} is defined by
\ba\label{TobitBINGARCHmodel}
X_t\ =\ \max\big\{0,\ \min\{N,\ X_t^*\}\big\}
\quad\text{with}\quad
X_t^*\ =\ \mathcal S_{\fnui}(M_t),
\ea
where~$\mathcal S_{\fnui}(\mu)$, $\mu\in\bbr$, again denotes a $\bbz$-valued random operator. An analogous construction might be used for adapting the Tobit INAR model \eqref{TobitINARp} to the case of bounded counts. Again, future research should study the properties and abilities of these models.

\section{Real Applications}
\label{Real Applications}
In what follows, we illustrate the application of the novel STINGARCH model by a couple of real-world data examples. In addition, these data examples serve for discussing some of the potential extensions sketched in Section~\ref{Possible Extensions}.

\begin{figure}[t]
	\center\small
	\includegraphics[scale=0.62, viewport=15 15 415 235, clip=]{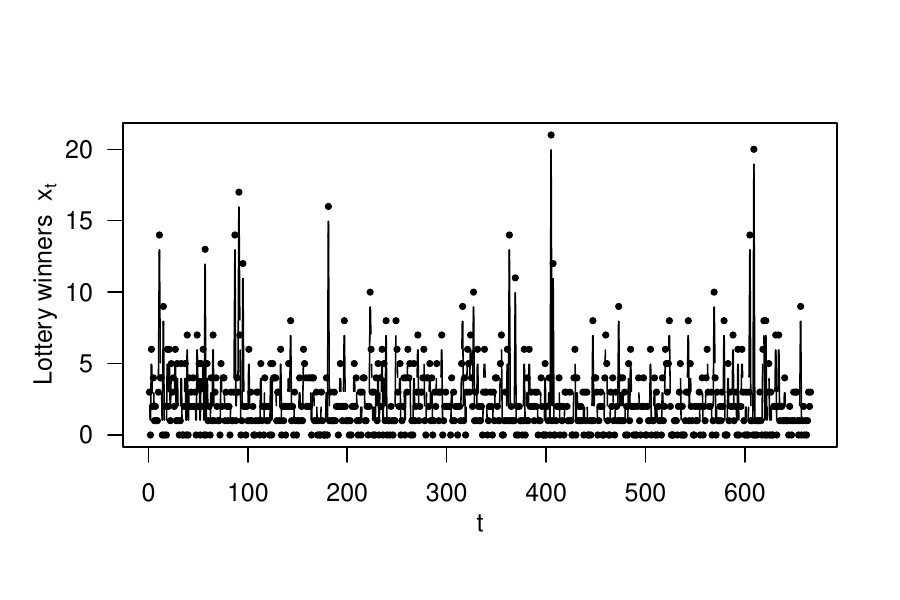}\qquad
	\includegraphics[scale=0.62, viewport=0 15 190 240, clip=]{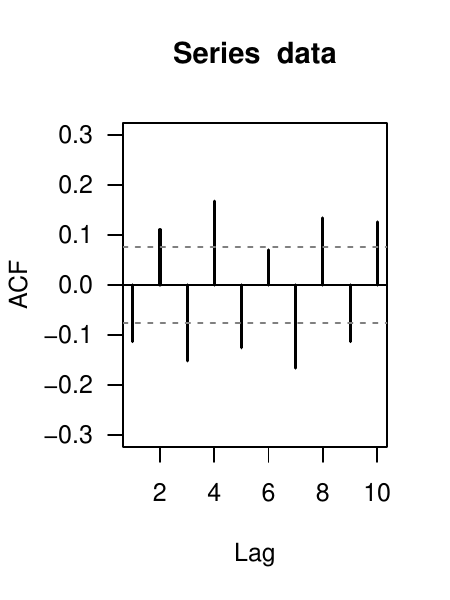}
	\caption{Plot of lottery winners $x_1,\ldots,x_{666}$ and their sample ACF $\hat{\rho}(k)$, see Section~\ref{Lottery Winners Data}.}
	\label{figLottery}
\end{figure}

\subsection{Lottery Winners Data}
\label{Lottery Winners Data}
The main lottery game in Germany is ``Lotto 6 aus 49'', where 6 out of 49 numbers (plus one bonus number) are randomly drawn twice a week (on Wednesdays and Saturdays). Information on past lottery draws is available on \url{https://www.westlotto.de/lotto-6aus49/gewinnzahlen/gewinnzahlen.html}, from which we obtained the number of lottery wins in the second prize category (``6 numbers'') for each lottery draw between Saturday, May 4, 2013, and Wednesday, September 18, 2019. The resulting count time series of length $n=666$ is plotted in Figure~\ref{figLottery}. The corresponding sample ACF shows an alternating pattern of positive and negative ACF values, which decrease only slowly with increasing time lag~$k$. Because of this kind of a ``long memory'', the STINGARCH$(1,1)$ model appears to be a plausible candidate for these data. Since it is not immediately clear that we are dealing with linear time series data here, we consider estimation scenario 2 (recall Section~\ref{Estimation}), \ie the parameter~$\delta$ is included into ML estimation.
The resulting ML estimates are shown in the first block of Table~\ref{tabLottery}. $\hat\delta$ takes a rather large value (similar for the remaining candidate models), which indicates a non-linear dependence structure (recall Figure~\ref{figLottery}). The strong value for $\hat\beta_1$, in turn, appears plausible in view of the apparent ``long memory''.
Since $|\hat\beta_1|$ is rather close to~1, we also fitted the STINGARCH$(1,2)$ model to the data. Its estimates are more balanced, but Akaike's and  Bayesian information criterion (AIC and BIC, respectively) are worse.
We also considered the Poisson softplus INGARCH$(1,1)$ (Poi-spINGARCH$(1,1)$) model of \citet{weissetal22} with $c=1$ as a further competitor (which is discussed in some more detail later in Section~\ref{Chemical Process Data}), but it clearly performs worse.
Computing the standardized Pearson residuals \citep[Sect.\ 2.4]{weiss18} for checking the adequacy of the ML-fitted models, their sample ACF values in Table~\ref{tabLotteryPearson} are not significant, indicating that both fitted models capture the ACF structure very well. Also their means are very close to zero, \ie the fitted models well describe the mean of the data (sample mean 2.443).
The variances of the residuals in Table~\ref{tabLotteryPearson} are only a little larger than one, and this slight exceedance is caused by a few extreme counts in Figure~\ref{figLottery}: excluding, for example, the residuals corresponding to $x_{405}=21$ and $x_{609}=20$, the variances reduce to 1.180 and 1.215. Thus, also the dispersion of the data appears to be covered quite well by the fitted models.

\begin{table}[t]
	\center\footnotesize
	\caption{ML estimation for lottery winners data: estimates and approximate standard errors (SEs), information criteria.}
	\label{tabLottery}
	
	\smallskip
	\resizebox{\linewidth}{!}{
	\begin{tabular}{@{}l|c@{\ }lc@{\ }lc@{\ }lc@{\ }lc@{\ }lll@{}}
	\toprule
		Model & $\hat\delta$ & \multicolumn{1}{c}{SE} & $\hat\alpha_0$ & \multicolumn{1}{c}{SE} & $\hat\alpha_1$ & \multicolumn{1}{c}{SE} & $\hat\beta_1$ & \multicolumn{1}{c}{SE} & $\hat\beta_2$ & \multicolumn{1}{c}{SE} & AIC & BIC \\
		\midrule
STINGARCH$(1,1)$ & 3.619 & (0.471) & 4.442 & (0.197) & -0.008 & (0.008) & -0.990 & (0.007) &  &  & 2782.7 & 2800.7 \\
STINGARCH$(1,2)$ & 3.595 & (0.483) & 1.630 & (0.617) & -0.044 & (0.017) & -0.316 & (0.137) & 0.642 & (0.136) & 2800.1 & 2822.6 \\
Poi-spINGARCH$(1,1)$ & && 5.023 & (0.118) & -0.016 & (0.011) & -1.099 & (0.012) & && 2923.3 & 2936.8 \\
\bottomrule
\multicolumn{13}{@{}c@{}}{}\\
\toprule
		Model & $\hat\delta$ & \multicolumn{1}{c}{SE} & $\hat\alpha_0$ & \multicolumn{1}{c}{SE} & $\hat\gamma_1$ & \multicolumn{1}{c}{SE} &  &  &  &  & AIC&BIC\\
		\midrule
ST-regression & 3.458 & (0.124) & 1.239 & (0.185) & 2.031 & (0.471) &  &  &  &  & 2783.3 & 2796.8 \\
\bottomrule
	\end{tabular}
	}

\vspace{3ex}
	\caption{Model diagnostics by Pearson residuals for lottery winners data.}
	\label{tabLotteryPearson}
	
	\smallskip
	\begin{tabular}{@{}l|ccccccc}
	\toprule
		Model & Mean & Variance & ACF(1) & ACF(2) & ACF(3) & ACF(4) & ACF(5) \\
		\midrule
STINGARCH$(1,1)$ & -0.007 & 1.351 & 0.012 & 0.008 & -0.043 & 0.042 & -0.009 \\
STINGARCH$(1,2)$ & -0.021 & 1.368 & 0.025 & 0.006 & -0.036 & 0.050 & -0.005 \\
ST-regression  & -0.019 & 1.358 & -0.009 & 0.010 & -0.060 & 0.057 & -0.026 \\
\bottomrule
	\end{tabular}
\end{table}

\smallskip
Revisiting the ACF in Figure~\ref{figLottery}, another explanation for the alternating pattern could be periodicity with period~2, which indeed appears plausible in view of the draws on Wednesdays and Saturdays. Hence, as a further candidate for modelling the lottery winners data, let us use the weekday information as a covariate, say, $z_t=1$ ($=0$) if the $t$th lottery draw is on Saturday (Wednesday). As already mentioned in Section~\ref{Possible Extensions}, covariates $z_{t,1},\ldots,z_{t,r}$ are easily included into the model by just adding the sum $\sum_{k=1}^r \gamma_k\,z_{t,k}$ on the right-hand side of Equation \eqref{LinCondMean}. For the ML-fit including the covariate term $+\gamma_1\,z_{t}$ referring to the weekday, the autoregressive parameters are not significant anymore, and the model can be reduced to the equation $M_t = \alpha_0+\gamma_1\,z_{t}$. Its estimation results (including~$\hat\delta$) are shown in the second block (``ST-regression'') of Table~\ref{tabLottery}. The positive value for $\hat{\gamma}_1$ implies that the event ``6~numbers'' is more likely on Saturdays than on Wednesdays, which can be explained by the fact that more people participate in Saturday's lottery. While the AIC is slightly worse than for the STINGARCH$(1,1)$ model, the BIC is better. The checks for model adequacy based on the Pearson residuals, see Table~\ref{tabLotteryPearson}, indicate a similarly well performance: their mean is close to zero, their variance (without $x_{405},x_{609}$: 1.209) is close to one, and their ACF is not significant. Hence, the user may decide between the candidate models, \eg based on their interpretability.

\begin{figure}[t]
	\center\small
	\includegraphics[scale=0.62, viewport=15 15 415 235, clip=]{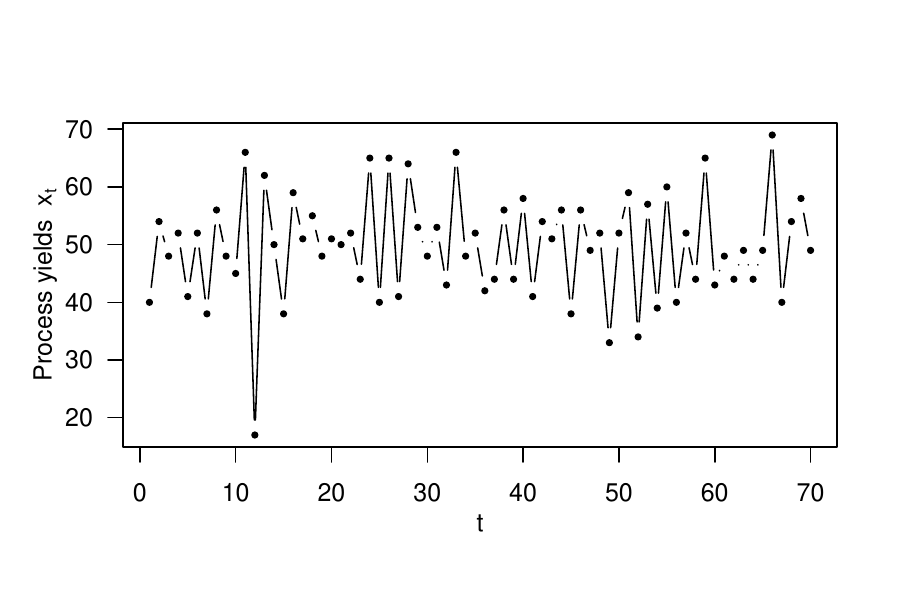}\qquad
	\includegraphics[scale=0.62, viewport=0 15 190 240, clip=]{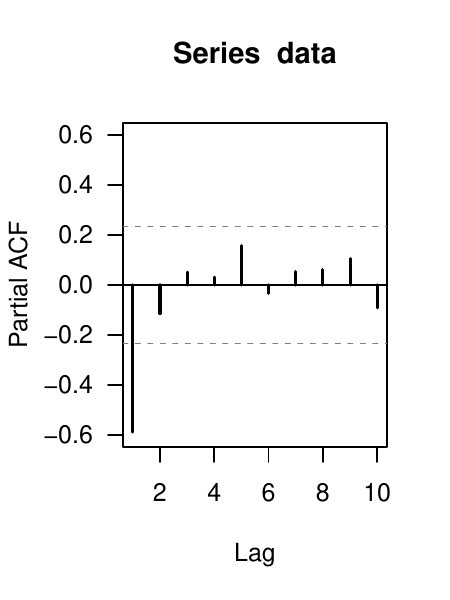}
	\caption{Plot of process yields $x_1,\ldots,x_{70}$ and their sample PACF $\hat{\rho}_{\textup{p}}(k)$, see Section~\ref{Chemical Process Data}.}
	\label{figChemical}
\end{figure}

\subsection{Chemical Process Data}
\label{Chemical Process Data}
As our second example, let us revisit the chemical process data in Appendix~A.3 of \citet[``Time series~4.1'']{odonovan83}, which consist of $n=70$ consecutive yields from a batch chemical process. The time series plot in Figure~\ref{figChemical} shows a ``spiky behaviour''. Because of the apparent AR$(1)$-like PACF with strongly negative lag-1 autocorrelation, $\hat{\rho}(1)= -0.588$, these data were considered by \citet[Section~5.2]{weissetal22} as an illustrative example for their Poi-spINARCH$(1)$ model. As explained in Section~\ref{Introduction}, our STINARCH$(1)$ constitutes a natural competitor to that model, so we shall now compare their model fits. In addition, recall that our Tobit approach is not limited to unbounded INGARCH-like counts, as it is the case for the softplus INGARCH model, but can also be adapted to different model families, see Section~\ref{Possible Extensions}. Also this topic shall be made a subject of discussion in this section.

\begin{table}[t]
	\center\footnotesize
	\caption{ML estimation for process yields data: estimates and approximate standard errors (SEs), information criteria.}
	\label{tabChemical}
	
	\smallskip
	\begin{tabular}{@{}ll|c@{\ }lc@{\ }lll@{}}
	\toprule
		Model & $c$ or $\delta$ & $\hat\alpha_0$ & \multicolumn{1}{c}{SE} & $\hat\alpha_1$ & \multicolumn{1}{c}{SE}  & AIC & BIC \\
		\midrule
Poi-spINARCH$(1)$ & 1 & 79.783 & (4.820) & -0.603 & (0.094) & 486.94 & 491.44 \\
STINARCH$(1)$ & 0.25 & 79.767 & (4.833) & -0.602 & (0.094) & 486.81 & 491.31 \\
Poi-TINARS$(1)$ & --- & 73.792 & (6.066) & -0.482 & (0.121) & 489.18 & 493.68 \\
\bottomrule
	\end{tabular}

\vspace{3ex}
	\caption{Model diagnostics by Pearson residuals for process yields data.}
	\label{tabChemicalPearson}
	
	\smallskip
	\begin{tabular}{@{}l|ccccccc}
	\toprule
		Model & Mean & Variance & ACF(1) & ACF(2) & ACF(3) & ACF(4) & ACF(5) \\
		\midrule
Poi-spINARCH$(1)$ & 0.000 & 1.142 & -0.064 & -0.054 & 0.119 & 0.044 & 0.082 \\
STINARCH$(1)$ & 0.000 & 1.136 & -0.065 & -0.054 & 0.118 & 0.044 & 0.082 \\
\bottomrule
	\end{tabular}
\end{table}

\smallskip
Comparing the ML-estimation results for the Poi-spINARCH$(1)$ and  STINARCH$(1)$ models, see rows 1--2 in Table~\ref{tabChemical}, we recognize a nearly identical performance (with slightly better AIC and BIC for STINARCH$(1)$). This also applies to the stochastic properties of both model fits. Regarding the Pearson residuals in Table~\ref{tabChemicalPearson}, both have a mean about zero, \ie they nearly perfectly capture the data's mean (sample mean 49.686).
The Pearson residuals' variances only slightly exceed one, see Table~\ref{tabChemicalPearson}, so both models only slightly fall below the sample dispersion~1.706.
Also the data's autocorrelation structure is described quite well by both model fits, because none of the Pearson residuals' ACF value is significantly different from zero.
Note that the STINARCH$(1)$'s CLADE are given by $\check{\alpha}_0=72.892$ and $\check{\alpha}_1=-0.461$, whereas CLS leads to $\breve{\alpha}_0=79.071$ and $\breve{\alpha}_1=-0.588$. Hence, like in the simulation study of Section~\ref{Results from Simulation Study}, the CLS estimates are more close to the MLE. Finally, like the Poi-spINARCH$(1)$ model's tuning parameter~$c$ was specified as~1, to guarantee conditional linearity in close approximation, we specified the STINARCH$(1)$'s tuning parameter as $\delta=0.25$ (as recommended in Section~\ref{On Moments and Approximate Linearity}; estimation scenario~1). Actually, because of the small sample size $n=70$, the estimation of~$\delta$ does not work reliably here because of huge SEs. But we noted that a variation of~$\delta$ had only negligible effects on the estimation results anyway.

\smallskip
In the previous paragraph, we recognized that Poi-spINARCH$(1)$ and  STINARCH$(1)$ lead to nearly equally well model fits. But as shown in Section~\ref{Possible Extensions}, the Tobit approach is more flexible and might be applied to different model families as well. As we are concerned with AR$(1)$-like unbounded counts, also the Tobit INARS$(1)$ model \eqref{TobitINARp} would be a plausible further candidate for the chemical process data. Thus, let us also fit the Tobit INARS$(1)$ model $X_t\, =\, \max\big\{0,\, \alpha_1\odot X_{t-1} + \epsilon_t\big\}$ with \iid\ $\poi(\alpha_0)$-innovations (Poi-TINARS$(1)$) to the data. Its transition probabilities, as required for likelihood computation, are
$$
\begin{array}{rl}
P(X_t=x>0\ |\ X_{t-1})\ =&
\sum_{j=0}^{X_{t-1}} P\Big(\bin\big(X_{t-1}, |\alpha_1|\big) = j\Big)\cdot P\Big(\epsilon_t=x-\textup{sgn}(\alpha_1)\,j\Big),
\\[1ex]
P(X_t=0\ |\ X_{t-1})\ =&
\sum_{j=0}^{X_{t-1}} P\Big(\bin\big(X_{t-1}, |\alpha_1|\big) = j\big)\cdot P\Big(\epsilon_t\leq -\textup{sgn}(\alpha_1)\,j\Big),
\end{array}
$$
where ``$\bin(X_{t-1}, |\alpha_1|)$'' is used as an abbreviation for a binomial random variable. But as can be seen from the third row in Table~\ref{tabChemical}, the fitted Poi-TINARS$(1)$ model performs worst. It does not capture the ACF to full extent ($\hat\alpha_1=-0.482$ vs.\ $\hat{\rho}(1)= -0.588$), and the fitted models dispersion ratio 2.257 is too large. So taking all results together, the chemical process data are best explained by the STINARCH$(1)$ model.

\begin{figure}[t]
	\center\small
	\includegraphics[scale=0.62, viewport=15 15 415 235, clip=]{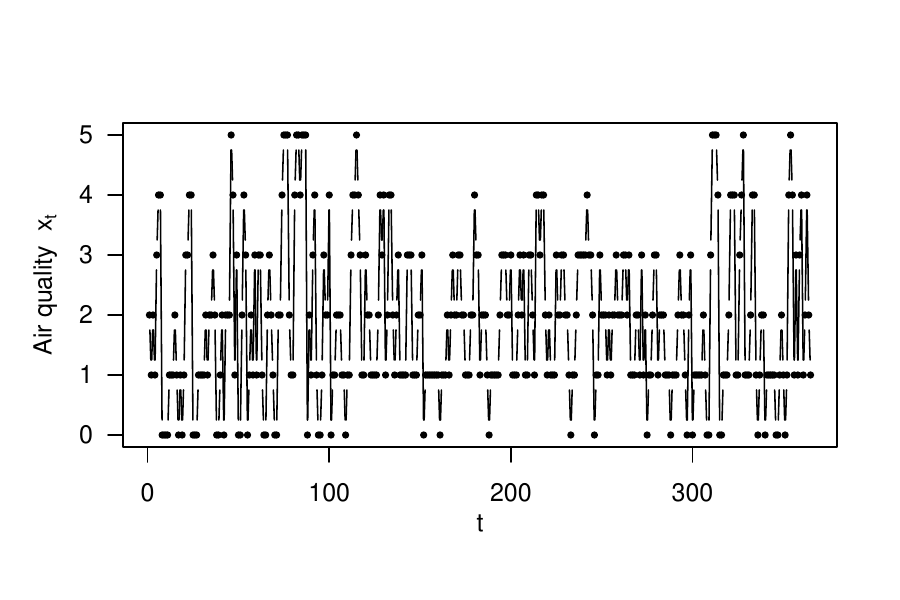}\qquad
	\includegraphics[scale=0.62, viewport=0 15 190 240, clip=]{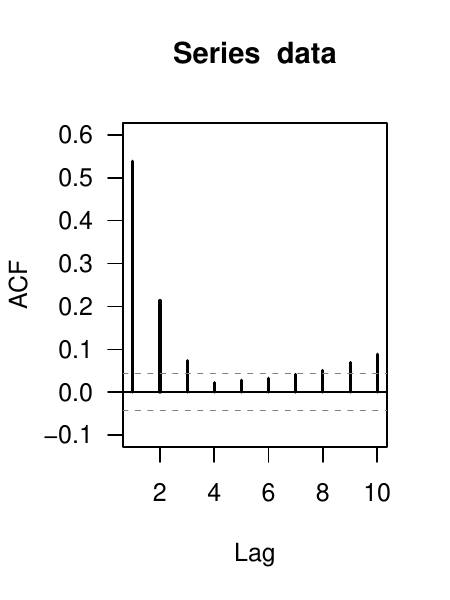}
	\caption{Plot of first 365 air quality counts, and sample ACF $\hat{\rho}(k)$ of full data, see Section~\ref{Air Quality Data}.}
	\label{figAir}
\end{figure}

\subsection{Air Quality Data}
\label{Air Quality Data}
Also the third example serves to demonstrate the great flexibility of the Tobit approach. For this purpose, let us revisit one of the time series discussed by \citet{liu22}, namely the daily air quality level (Dec.\ 2013--July 2019, so length $n=2\,068$) in Beijing. While air quality is actually measured on an ordinal scale (from $s_0=\text{``excellent''}$ to $s_5=\text{``severely polluted''}$), \citet{liu22} used a ``rank-count approach'' for modelling these data, \ie the ordinal~$Y_{t}$ at time~$t$ is substituted by $s_{X_{t}}$ with the count $X_{t}$ having the upper bound $N=5$. For the sake of readability, the plot in Figure~\ref{figAir} shows only the first 365 observations, as well as the sample ACF $\hat{\rho}(k)$ of the full data. \citet{liu22} fitted an (exactly linear) INGARCH$(1,1)$ model to the data, referred to as ``ZOBPAR$(1,1)$'', where the conditional distribution is a truncated Poisson distribution with additional one-inflation. Because of being exactly linear, the ZOBPAR's model parameters are not allowed to become negative. However, looking at the sample ACF in Figure~\ref{figAir}, we observe a rather strong decline after lag~1, and this might be better explained if negative parameter values are permitted. Thus, we use the Skellam Tobit BINGARCH$(1,1)$ (STBINGARCH$(1,1)$) model \eqref{TobitBINGARCHmodel} with $N=5$ from Section~\ref{Possible Extensions} as a competitor, which we also equip with a further parameter $\kappa\in [0;1)$ for capturing one-inflation:
$$
P(X_t=x\ |\ X_{t-1},\ldots)\ =\ \kappa\cdot\indfkt_{\{1\}}(x)\ +\ (1-\kappa)\cdot\left\{\begin{array}{ll}
P\big(\mathcal{S}_\delta(M_t)\leq 0\big) & \text{if } x=0,\\[.5ex]
P\big(\mathcal{S}_\delta(M_t)=x\big) & \text{if } 1\leq x\leq N-1,\\[.5ex]
P\big(\mathcal{S}_\delta(M_t)\geq N\big) & \text{if } x=N.
\end{array}\right.
$$
When specifying~$\delta$ in our data applications, we routinely tried out candidate models with different levels~$\delta$ of linearity. In this case, we recognized that we can further improve upon our default choice $\delta=0.25$ by choosing the tuning parameter very close to~0; we used $\delta=0.01$.
The results of ML estimation are summarized in Table~\ref{tabAir}.

\begin{table}[t]
	\center\small
	\caption{ML estimation for air quality data: estimates and approximate standard errors (SEs), information criteria.}
	\label{tabAir}
	
	\smallskip
	\resizebox{\linewidth}{!}{
	\begin{tabular}{@{}ll|c@{\ }lc@{\ }lc@{\ }lc@{\ }lll@{}}
	\toprule
		Model & $\delta$ & $\hat\alpha_0$ & \multicolumn{1}{c}{SE} & $\hat\alpha_1$ & \multicolumn{1}{c}{SE} & $\hat\beta_1$ & \multicolumn{1}{c}{SE} & $\kappa$ & \multicolumn{1}{c}{SE}  & AIC & BIC \\
		\midrule
ZOBPAR$(1,1)$ & --- & 0.592 & (0.059) & 0.756 & (0.038) & 0.000 & (0.040) & 0.128 & (0.015) & 5721.6 & 5744.1 \\
STBINGARCH$(1,1)$ & 0.01 & 0.787 & (0.067) & 0.699 & (0.030) & -0.127 & (0.037) & 0.118 & (0.015) & 5745.5 & 5768.0 \\
\bottomrule
	\end{tabular}
	}

\vspace{3ex}
	\caption{Model diagnostics by Pearson residuals for air quality data.}
	\label{tabAirPearson}
	
	\smallskip
	\begin{tabular}{@{}l|ccccccc}
	\toprule
		Model & Mean & Variance & ACF(1) & ACF(2) & ACF(3) & ACF(4) & ACF(5) \\
		\midrule
ZOBPAR$(1,1)$ & 0.010 & 0.801 & 0.060 & -0.059 & -0.028 & -0.023 & 0.014 \\
STBINGARCH$(1,1)$ & -0.034 & 0.752 & 0.018 & -0.087 & -0.043 & -0.031 & 0.011 \\
\bottomrule
	\end{tabular}
\end{table}

\smallskip
If solely looking at the information criteria, then Table~\ref{tabAir} clearly votes for the ZOBPAR$(1,1)$ model of \citet{liu22}. But in fact, a more in-depth analysis is advisable. It can be seen that the $\beta_1$-estimate of the ZOBPAR model falls on the lower bound~0 of its admissible range, whereas STBINGARCH leads to a significantly negative~$\hat\beta_1$. If comparing the ACFs of the fitted models to the sample ACF, namely $0.521, 0.272, 0.144, \ldots$ (ZOBPAR) vs.\ $0.527, 0.228, 0.100, \ldots$ (STBINGARCH) vs.\ $0.539, 0.214, 0.073, \ldots$ (data), we recognize that STBINGARCH better explains the strong decline after lag~1.
This is also confirmed by an analysis of the Pearson residuals' ACF in Table~\ref{tabAirPearson}, where the ZOBPAR model has significant values at lags~1 and~2, and STBINGARCH only at lag~2. Furthermore, both models lead to a Pearson residuals' variance below one this time,
\ie both models include too much dispersion, and now STBINGARCH does worse. The reason becomes clear by comparing the full marginal distributions $p(0),\ldots,p(5)$:
$$
\begin{array}{rlllllll}
 & 0.220, & 0.375, & 0.183, & 0.117, & 0.068, & 0.037 & \text{(ZOBPAR)} \\
\text{vs.} & 0.229, & 0.374, & 0.182, & 0.107, & 0.057, & 0.051 & \text{(STBINGARCH)} \\
\text{vs.} & 0.174, & 0.383, & 0.222, & 0.131, & 0.065, & 0.025 & \text{(data).}
\end{array}
$$
Both models over-estimate the zero probability at the cost of under-estimating some central probabilities, but STBINGARCH also over-estimates the probability for reaching the upper bound $N=5$. Thus, the STBINGARCH's disadvantage regarding the information criteria seems to be caused by this deviation of the marginal distribution, whereas it performs better regarding the autocorrelation structure.

\section{Conclusions}
\label{Conclusions}
To give an alternative solution to the log and softplus approaches for achieving negative ACF values within the class of INGARCH models, the Tobit approach is proposed, and the Skellam--Tobit INGARCH model for unbounded counts is studied in detail. Several extensions of this approach to other situations are discussed as well. Three real-data examples are analyzed to illustrate the usefulness of the new approach.

There are some suggestions for possible future research. First, one could work out the TINGARCH model based on the generalized Poisson difference distribution (recall Sections~\ref{Possible Extensions})
or other distributions. In particular, closed-form expressions for distribution and moments of the corresponding censored random variable $X^*\,\indfkt_{\bbn}(X^*)$ should be derived.
Second, more studies regarding the Tobit INAR models for unbounded or bounded counts mentioned in Sections~\ref{Possible Extensions} are needed.
Third, similar to the arguments in \citet{weissetal22}, we can apply the new approach to many real datasets and find some causal relationship like \citet{chen18}, who investigated the relation between human influenza cases and air pollution.

\subsubsection*{Acknowledgements}
The authors  thank Dr.\ Tobias M\"oller (Helmut Schmidt University, Hamburg) for the collection and preparation of the lottery winners data used in Section~\ref{Lottery Winners Data}.
Zhu's work is supported by National Natural Science Foundation of China (No. 12271206),  and Science and Technology Research Planning Project of Jilin Provincial Department of Education (No. JJKH20231122KJ).

\appendix\small
\numberwithin{equation}{section}
\numberwithin{table}{section}

\section{Appendices}
\label{Appendices}

\subsection{About INGARCH models}
\label{INGARCH}
For INGARCH$(p,q)$ models, the count~$X_t$ being generated at time~$t$ has the conditional mean~$M_t$ satisfying the linear recursive scheme \eqref{INGARCHmodel} with $a_0>0$ and $a_1,\ldots,a_{p},b_1,\ldots,b_{q}\geq 0$. The default choice is a conditional Poisson distribution, \ie $X_t$, conditioned on $X_{t-1},\ldots$, is $\poi(M_t)$-distributed.
Using the linearity of the conditional mean, the unconditional mean equals \citep{ferland06}
\ba
\label{INGARCHmean}
\mu\ =\ \frac{a_0}{1-\sum_{i=1}^p a_i-\sum_{j=1}^q b_j},
\ea
and variance and autocovariances satisfy a set of Yule--Walker equations \citep{weiss09}:
\ba
\label{INGARCHyw}
\begin{array}{@{}rl}
\gamma(0)\ =& \mu\ +\ \gamma_M(0),
\qquad
\gamma_M(0)\ =\ \sum\limits_{i=1}^{p} a_i\, \gamma(i)\ +\ \sum\limits_{j=1}^{q} b_j\, \gamma_M(j),\\[3ex]
\gamma(k)\ =& \sum\limits_{i=1}^{p} a_i\, \gamma(|k-i|)\ +\ \sum\limits_{j=1}^{\min{\{k-1,q\}}} b_j\, \gamma(k-j)\ +\ \sum\limits_{j=k}^{q} b_j\, \gamma_M(j-k),\\[3ex]
\gamma_M(k)\ =& \sum\limits_{i=1}^{\min{\{k,p\}}} a_i\, \gamma_M(|k-i|)\ +\ \sum\limits_{i=k+1}^{p} a_i\, \gamma(i-k)\ +\ \sum\limits_{j=1}^{q} b_j\, \gamma_M(|k-j|),\\
\end{array}
\ea
for $k\geq 1$, where $\gamma(h) := Cov[X_t, X_{t-h}]$ and $\gamma_{M}(h):=Cov[M_t, M_{t-h}]$.
Because of the conditional Poisson distribution, the conditional dispersion ratio $V(X_t|\mathcal F_{t-1}) \big/ E(X_t|\mathcal F_{t-1})=1$ constantly (the unconditional distribution exhibits overdispersion, \ie $\sigma^2/\mu>1$). To allow for a more flexible dispersion structure, several extensions of the basic Poisson INGARCH model have been developed in the literature, see \citet{weiss18} for a survey. For this article, especially the extension by \citet{xu12} is relevant, who defined the DINGARCH model by introducing an additional dispersion parameter $\eta\geq 1$ into the conditional variance:
\ba
\label{DINGARCH}
V(X_t|\mathcal F_{t-1})\ =\ \eta\, E(X_t|\mathcal F_{t-1}).
\ea
So the Poisson INGARCH model corresponds to the choice $\eta=1$, while specific examples for $\eta>1$ are given by the negative-binomial INGARCH model of \citet{xu12} and the generalized-Poisson INGARCH model of \citet{zhu12}. As a result, the mean and ACF of the DINGARCH model are the same as for the Poisson INGARCH model, but the variance is inflated by the factor~$\eta$.

\subsection{Proof of Proposition~\ref{propSkellamPartmom}}
\label{Proof of Proposition propSkellamPartmom}
To derive \eqref{Skellam_partial}, we set $f(x)=\indfkt_{\bbn}(x)$ in \eqref{Skellam_Stein} and obtain
$$
\begin{array}{rl}
E\big(X^*\,\indfkt_{\bbn}(X^*)\big)\ =&
\lambda_1\,E\big(\indfkt_{\bbn}(X^*+1)\big)\ -\
\lambda_2\,E\big(\indfkt_{\bbn}(X^*-1)\big)
\\[1ex]
=&
\lambda_1\,P(X^*\geq 0)\ -\
\lambda_2\,P(X^*\geq 2)
\\[1ex]
=&
\mu\,P(X^*\geq 0)\ +\
\lambda_2\,\big(P(X^*=0) + P(X^*=1)\big).
\end{array}
$$
Next, let us turn to the second-order partial moment \eqref{Skellam_pmom2}.
In a first step, we apply \eqref{Skellam_Stein} with $f(x)=x\,\indfkt_{\bbn}(x)$, leading to
$$
\begin{array}{@{}l}
E\big((X^*)^2\,\indfkt_{\bbn}(X^*)\big)\ =\
\lambda_1\,E\big((X^*+1)\,\indfkt_{\bbn}(X^*+1)\big)\ -\
\lambda_2\,E\big((X^*-1)\,\indfkt_{\bbn}(X^*-1)\big)
\\[1ex]
\quad =\
\lambda_1\,E\big(X^*\,\indfkt_{\bbn}(X^*+1)\big)\ -\
\lambda_2\,E\big(X^*\,\indfkt_{\bbn}(X^*-1)\big)
\ +\
\lambda_1\,P(X^*\geq 0)\ +\
\lambda_2\,P(X^*\geq 2).
\end{array}
$$
In a second step, we apply \eqref{Skellam_Stein} with $f(x)=\indfkt_{\bbn}(x+1)$ or $f(x)=\indfkt_{\bbn}(x-1)$, respectively, leading to
$$
\begin{array}{rl}
E\big(X^*\,\indfkt_{\bbn}(X^*+1)\big)\ =&
\lambda_1\,P(X^*\geq -1)\ -\
\lambda_2\,P(X^*\geq 1)
\\[.5ex]
\ =&
\lambda_1\,P(X^*= -1) + \lambda_1\,P(X^*=0)\ +\ (\lambda_1-\lambda_2)\,P(X^*\geq 1)
\\[.5ex]
\ \overset{\eqref{skellam2}}{=}&
\lambda_2\,P(X^*= 1) + \lambda_1\,P(X^*=0)\ +\ (\lambda_1-\lambda_2)\,P(X^*\geq 1),\\[1ex]
E\big(X^*\,\indfkt_{\bbn}(X^*-1)\big)\ =&
\lambda_1\,P(X^*\geq 1)\ -\
\lambda_2\,P(X^*\geq 3)
\\[.5ex]
=&
\lambda_2\,P(X^*=1)\ +\ \lambda_2\,P(X^*=2)\ +\ (\lambda_1-\lambda_2)\,P(X^*\geq 1)
\\[.5ex]
\ \overset{\eqref{skellam3}}{=}&
(\lambda_2-1)\,P(X^*=1)\ +\ \lambda_1\,P(X^*=0)\ +\ (\lambda_1-\lambda_2)\,P(X^*\geq 1).
\end{array}
$$
So altogether, we get
$$
\begin{array}{@{}l}
E\big((X^*)^2\,\indfkt_{\bbn}(X^*)\big)\ =\ \lambda_1\,P(X^*\geq 0)\ +\
\lambda_2\,P(X^*\geq 2)\ +\ (\lambda_1-\lambda_2)^2\,P(X^*\geq 1)
\\[.5ex]
\qquad +\
\lambda_1\lambda_2\,P(X^*= 1) + \lambda_1^2\,P(X^*=0)
\ -\
\lambda_2\,(\lambda_2-1)\,P(X^*=1) - \lambda_1\lambda_2\,P(X^*=0)
\\[1ex]
\quad =\ \lambda_1\,P(X^*\geq 0)\ +\
\lambda_2\,P(X^*\geq 1)\ +\ (\lambda_1-\lambda_2)^2\,P(X^*\geq 1)
\\[.5ex]
\qquad +\
\lambda_2\,(\lambda_1-\lambda_2)\,P(X^*= 1)
\ +\
\lambda_1\,(\lambda_1-\lambda_2)\,P(X^*=0)
\\[1ex]
\quad =\ (\sigma^2+\mu^2)\,P(X^*\geq 1)\ +\
\lambda_2\,\mu\,P(X^*= 1)
\ +\
\lambda_1\,(1+\mu)\,P(X^*=0).
\end{array}
$$
This completes the proof of Proposition~\ref{propSkellamPartmom}.

\subsection{Proof of Theorem~\ref{thmExistence}}
\label{Proof of Theorem thmExistence}
We just need to show the existence and stationarity of the process $\{X_t^\ast\}$.
\citet{carallo20} considered a process similar to $\{X_t^\ast\}$, see their equation (5), where they assume that $\alpha_1,\ldots,\alpha_p,\beta_1,\ldots,\beta_q$ are non-negative but $\alpha_0\in\bbr$.
The non-negativity restriction on the parameters is redundant, as it does not affect the proof of \citet{carallo20}. 

For $M_t$ defined in our \eqref{LinCondMean}, we have
\begin{align*}
\textstyle
E(X_t^\ast|\mathcal{F}_{t-1})=M_t\ =\ \alpha_0+\sum\limits_{i=1}^p\alpha_iX_{t-i}^\ast\indfkt_{\bbn}(X_{t-i}^\ast)+\sum\limits_{j=1}^q\beta_jM_{t-j}.
\end{align*}
Based on the sign of $X_{t-i}^\ast$, $i=1,\ldots,p$, $M_t$ can have $2^p$ different cases.
We consider the special case $M_t=\alpha_0+\sum_{i=1}^p\alpha_iX_{t-i}^\ast+\sum_{j=1}^q\beta_jM_{t-j}$. Using arguments similar to those in Section 3.1 in \citet{carallo20}, we know that if
condition \eqref{st-cond} is satisfied, then the existence and stationarity of the process $\{X_t^\ast\}$ holds.
For remaining cases, where at least one of the summands in $\sum_{i=1}^p\alpha_iX_{t-i}^\ast$ vanishes, we can obtain similar conditions. But condition \eqref{st-cond} is the strictest, so it is a sufficient condition for all cases.

\subsection{Derivatives of Log-Likelihood Function (\ref{MLE})}
\label{Derivatives of Log-Likelihood Function}
In what follows, the derivatives of the log-likelihood function $L(\ftheta)=\sum_{t=1}^n l_t(\ftheta)$ according to \eqref{MLE} are discussed.
If $X_t>0$ and $M_t>0$, then
\begin{align*}
  \frac{\partial l_t(\ftheta)}{\partial\ftheta^*}&\ =\ \left(-1+\frac{X_t}{a_t}+\frac{\sqrt\delta}{\sqrt{a_t}}\,b_t\right)\frac{\partial M_t}{\partial\ftheta^*},
	\\
  \frac{\partial l_t(\ftheta)}{\partial\delta}&\ =\ -1+\frac{X_t}{2a_t}-\frac{X_t}{2\delta}+\frac{M_t+\delta}{\sqrt{\delta a_t}}\,b_t,
	\\
    \frac{\partial^2l_t(\ftheta)}{\partial\ftheta^*\partial\ftheta^{*\top}}&\ =\ \frac{1}{a_t}\left(-\frac{2X_t}{a_t}-\frac{\sqrt\delta}{\sqrt a_t}\, b_t+\delta c_t\right)\frac{\partial M_t}{\partial\ftheta^*}\,\frac{\partial M_t}{\partial\ftheta^{*\top}}
  +\left(-1+\frac{X_t}{a_t}+\frac{\sqrt\delta}{\sqrt{a_t}}\,b_t\right)\frac{\partial^2 M_t}{\partial\ftheta^*\partial\ftheta^{*\top}},
	\\
    \frac{\partial^2l_t(\ftheta)}{\partial^2\delta^2}&\ =\ -\frac{X_t}{2a_t^2}+\frac{X_t}{2\delta^2}-\frac{M_t^2}{(\delta a_t)^{3/2}}\,b_t+\frac{(M_t+\delta)^2}{\delta a_t}\,c_t,
		\\
    \frac{\partial^2l_t(\ftheta)}{\partial\ftheta^*\partial\delta}&\ =\ \frac{1}{a_t}\left(-\frac{X_t}{a_t}
    +\frac{M_t}{\sqrt{\delta a_t}}\,b_t
    +(M_t+\delta)\,c_t\right)\frac{\partial M_t}{\partial\ftheta^*},
\end{align*}
where
$$
a_t\ =\ 2M_t+\delta,\qquad
b_t\ =\ \frac{I_{X_t}'}{I_{X_t}},\qquad
c_t\ =\ \frac{I_{X_t}''I_{X_t}-(I_{X_t}')^2}{I_{X_t}^2}.
$$
Here, the functions $I_{X_t},I_{X_t}'$ and $I_{X_t}''$ omit their argument $\sqrt{\delta\,a_t}$ for brevity.
From \eqref{derbessel}, we can calculate the derivatives required for $b_t$ and $c_t$ as follows:
\begin{align*}
I_{X_t}'&\ =\ \frac{X_t}{\sqrt{\delta a_t}}\,I_{X_t}+I_{X_t+1},\\
I_{X_t}''&\ =\ \frac{1}{2}(I_{X_t-1}'+I_{X_t+1}')
\ =\ I_{X_t}+\frac{1}{2\sqrt{\delta a_t}}\,
\Big[(X_t-1)\,I_{X_t-1}-(X_t+1)\,I_{X_t+1}\Big].
\end{align*}
For $i,i'=1,\ldots,p,j,j'=1,\ldots,q$ and $i\neq i',j\neq j'$, we have
\begin{align*}
  \frac{\partial M_t}{\partial\alpha_0}&\ =\ 1+\sum_{l=1}^q\beta_l\,\frac{\partial M_{t-l}}{\partial\alpha_0},\quad
  \frac{\partial M_t}{\partial\alpha_i}\ =\ X_{t-i}+\sum_{l=1}^q\beta_l\,\frac{\partial M_{t-l}}{\partial\alpha_i},\quad
  \frac{\partial M_t}{\partial\beta_j}\ =\ M_{t-j}+\sum_{l=1}^q\beta_l\,\frac{\partial M_{t-l}}{\partial\beta_j},
	\\
  \frac{\partial^2M_t}{\partial\alpha_0^2}&\ =\ \sum_{l=1}^q\beta_l\,\frac{\partial^2M_{t-l}}{\partial\alpha_0^2},\qquad
  \frac{\partial^2M_t}{\partial\alpha_i^2}\ =\ \sum_{l=1}^q\beta_l\,\frac{\partial^2M_{t-l}}{\partial\alpha_i^2},\qquad
  \frac{\partial^2M_t}{\partial\alpha_0\partial\alpha_i}\ =\ \sum_{l=1}^q\beta_l\,\frac{\partial^2M_{t-l}}{\partial\alpha_0\partial\alpha_i},
	\\
   \frac{\partial^2M_t}{\partial\alpha_i\partial\alpha_{i'}}&\ =\ \sum_{l=1}^q\beta_l\,\frac{\partial^2M_{t-l}}{\partial\alpha_i\partial\alpha_{i'}},\qquad
  \frac{\partial^2M_t}{\partial\alpha_0\partial\beta_j}\ =\ \frac{\partial M_{t-j}}{\partial\alpha_0}+
  \sum_{l=1}^q\beta_l\,\frac{\partial^2M_{t-l}}{\partial\alpha_0\partial\beta_j},
  \\
  \frac{\partial^2M_t}{\partial\alpha_i\partial\beta_j}&\ =\ \frac{\partial M_{t-j}}{\partial\alpha_i}+
  \sum_{l=1}^q\beta_l\,\frac{\partial^2M_{t-l}}{\partial\alpha_i\partial\beta_j},\qquad
  \frac{\partial^2M_t}{\partial\beta_j^2}\ =\ 2\,\frac{\partial M_{t-j}}{\partial\beta_j}+
  \sum_{l=1}^q\beta_l\,\frac{\partial^2M_{t-l}}{\partial\beta_j^2},
	\\
  \frac{\partial^2M_t}{\partial\beta_j\partial\beta_{j'}}&\ =\ \frac{\partial M_{t-j}}{\partial\beta_{j'}}+
  \frac{\partial M_{t-j'}}{\partial\beta_{j}}+
  \sum_{l=1}^q\beta_l\,\frac{\partial^2M_{t-l}}{\partial\beta_j\partial\beta_{j'}}.
\end{align*}
Similarly, with an abuse of notations, if $X_t>0$ and $M_t<0$, then
\begin{align*}
  \frac{\partial l_t(\ftheta)}{\partial\ftheta^*}&\ =\ \left(1+\frac{X_t}{a_t}-\frac{\sqrt\delta}{\sqrt{a_t}}\,b_t\right)\frac{\partial M_t}{\partial\ftheta^*},
	\\
  \frac{\partial l_t(\ftheta)}{\partial\delta}&\ =\ -1-\frac{X_t}{2a_t}+\frac{X_t}{2\delta}-\frac{M_t-\delta}{\sqrt{\delta a_t}}\,b_t,
	\\
    \frac{\partial^2l_t(\ftheta)}{\partial\ftheta^*\partial\ftheta^{*\top}}&\ =\ \frac{1}{a_t}\left(\frac{2X_t}{a_t}+\frac{\sqrt\delta}{\sqrt a_t}\, b_t+\delta c_t\right)\frac{\partial M_t}{\partial\ftheta^*}\frac{\partial M_t}{\partial\ftheta^{*\top}}
 \ +\ \left(1+\frac{X_t}{a_t}-\frac{\sqrt\delta}{\sqrt{a_t}}\,b_t\right)\frac{\partial^2 M_t}{\partial\ftheta^*\partial\ftheta^{*\top}},
	\\
    \frac{\partial^2l_t(\ftheta)}{\partial^2\delta^2}&\ =\ \frac{X_t}{2a_t^2}-\frac{X_t}{2\delta^2}-\frac{M_t^2}{(\delta a_t)^{3/2}}\,b_t+\frac{(M_t-\delta)^2}{\delta a_t}\,c_t,
		\\
    \frac{\partial^2l_t(\ftheta)}{\partial\ftheta^*\partial\delta}&\ =\ \frac{1}{a_t}\left(-\frac{X_t}{a_t}
    +\frac{M_t}{\sqrt{\delta a_t}}\,b_t+(M_t-\delta)\,c_t\right)\frac{\partial M_t}{\partial\ftheta^*}.
\end{align*}
Now, $a_t=-2M_t+\delta$, and $b_t$ and $c_t$ are defined similarly as before. Here, the omitted variable in the functions $I_{X_t},I_{X_t}'$ and $I_{X_t}''$ in $b_t$ and $c_t$ is $\sqrt{\delta\,a_t}$.

If $X_t=0$, then the expression for $l_t(\ftheta)$ involves the sum of the conditional probabilities, but each term has been discussed above.

\clearpage

\section{Tables}
\label{Tables}

\begin{table}[h!]
\caption{Moment properties of STINARCH$(1)$ model, see Section~\ref{On Moments and Approximate Linearity}: true values (``tob'') vs.\ linear approximation (``lin''), mean close to~5.}
\label{tabMom5}

\smallskip
\resizebox{\linewidth}{!}{
\begin{tabular}{lrl|cc|cc|cc|cc|cc}
\toprule
 &&& \multicolumn{2}{c|}{$\mu$} & \multicolumn{2}{c|}{$\sigma^2/\mu$} & \multicolumn{2}{c|}{PACF$(1)$} & \multicolumn{2}{c|}{PACF$(2)$} & \multicolumn{2}{c}{PACF$(3)$} \\
$\alpha_0$ & $\alpha_1$ & $\delta$ & tob & lin & tob & lin & tob & lin & tob & lin & tob & lin \\
\midrule
8.75 & -0.75 & 1 & 5.044 & 5.000 & 2.303 & 2.710 & -0.698 & -0.750 & 0.024 & 0.000 & 0.007 & 0.000 \\
&& 0.5 & 5.032 & 5.000 & 2.195 & 2.501 & -0.707 & -0.750 & 0.023 & 0.000 & 0.007 & 0.000 \\
&& 0.25 & 5.027 & 5.000 & 2.139 & 2.394 & -0.712 & -0.750 & 0.022 & 0.000 & 0.007 & 0.000 \\
&& 0 & 5.021 & 5.000 & 2.082 & 2.286 & -0.717 & -0.750 & 0.021 & 0.000 & 0.007 & 0.000 \\
\midrule
7.5 & -0.50 & 1 & 5.009 & 5.000 & 1.557 & 1.581 & -0.493 & -0.500 & 0.001 & 0.000 & 0.000 & 0.000 \\
&& 0.5 & 5.004 & 5.000 & 1.448 & 1.459 & -0.496 & -0.500 & 0.001 & 0.000 & 0.000 & 0.000 \\
&& 0.25 & 5.002 & 5.000 & 1.391 & 1.397 & -0.498 & -0.500 & 0.000 & 0.000 & 0.000 & 0.000 \\
&& 0 & 5.000 & 5.000 & 1.332 & 1.333 & -0.499 & -0.500 & 0.000 & 0.000 & 0.000 & 0.000 \\
\midrule
6.25 & -0.25 & 1 & 5.005 & 5.000 & 1.262 & 1.265 & -0.248 & -0.250 & 0.000 & 0.000 & 0.000 & 0.000 \\
&& 0.5 & 5.002 & 5.000 & 1.166 & 1.167 & -0.249 & -0.250 & 0.000 & 0.000 & 0.000 & 0.000 \\
&& 0.25 & 5.001 & 5.000 & 1.117 & 1.117 & -0.250 & -0.250 & 0.000 & 0.000 & 0.000 & 0.000 \\
&& 0 & 5.000 & 5.000 & 1.067 & 1.067 & -0.250 & -0.250 & 0.000 & 0.000 & 0.000 & 0.000 \\
\midrule
3.75 & 0.25 & 1 & 5.009 & 5.000 & 1.263 & 1.265 & 0.249 & 0.250 & 0.000 & 0.000 & 0.000 & 0.000 \\
&& 0.5 & 5.003 & 5.000 & 1.166 & 1.167 & 0.249 & 0.250 & 0.000 & 0.000 & 0.000 & 0.000 \\
&& 0.25 & 5.002 & 5.000 & 1.117 & 1.117 & 0.250 & 0.250 & 0.000 & 0.000 & 0.000 & 0.000 \\
&& 0 & 5.000 & 5.000 & 1.067 & 1.067 & 0.250 & 0.250 & 0.000 & 0.000 & 0.000 & 0.000 \\
\midrule
2.5 & 0.50 & 1 & 5.019 & 5.000 & 1.567 & 1.581 & 0.497 & 0.500 & 0.000 & 0.000 & 0.000 & 0.000 \\
&& 0.5 & 5.008 & 5.000 & 1.453 & 1.459 & 0.499 & 0.500 & 0.000 & 0.000 & 0.000 & 0.000 \\
&& 0.25 & 5.004 & 5.000 & 1.394 & 1.397 & 0.499 & 0.500 & 0.000 & 0.000 & 0.000 & 0.000 \\
&& 0 & 5.000 & 5.000 & 1.333 & 1.333 & 0.500 & 0.500 & 0.000 & 0.000 & 0.000 & 0.000 \\
\midrule
1.25 & 0.75 & 1 & 5.091 & 5.000 & 2.606 & 2.710 & 0.744 & 0.750 & 0.000 & 0.000 & 0.000 & 0.000 \\
&& 0.5 & 5.042 & 5.000 & 2.453 & 2.501 & 0.747 & 0.750 & 0.000 & 0.000 & 0.000 & 0.000 \\
&& 0.25 & 5.020 & 5.000 & 2.372 & 2.394 & 0.748 & 0.750 & 0.000 & 0.000 & 0.000 & 0.000 \\
&& 0 & 5.000 & 5.000 & 2.286 & 2.286 & 0.750 & 0.750 & 0.000 & 0.000 & 0.000 & 0.000 \\
\bottomrule
\end{tabular}}
\end{table}

\clearpage

\begin{table}[h]
\caption{Moment properties of STINARCH$(1)$ model, see Section~\ref{On Moments and Approximate Linearity}: true values (``tob'') vs.\ linear approximation (``lin''), mean close to~10.}
\label{tabMom10}

\smallskip
\resizebox{\linewidth}{!}{
\begin{tabular}{lrl|cc|cc|cc|cc|cc}
\toprule
 &&& \multicolumn{2}{c|}{$\mu$} & \multicolumn{2}{c|}{$\sigma^2/\mu$} & \multicolumn{2}{c|}{PACF$(1)$} & \multicolumn{2}{c|}{PACF$(2)$} & \multicolumn{2}{c}{PACF$(3)$} \\
$\alpha_0$ & $\alpha_1$ & $\delta$ & tob & lin & tob & lin & tob & lin & tob & lin & tob & lin \\
\midrule
17.5 & -0.75 & 1 & 10.008 & 10.000 & 2.438 & 2.514 & -0.741 & -0.750 & 0.005 & 0.000 & 0.002 & 0.000 \\
&& 0.5 & 10.006 & 10.000 & 2.344 & 2.400 & -0.743 & -0.750 & 0.004 & 0.000 & 0.002 & 0.000 \\
&& 0.25 & 10.005 & 10.000 & 2.297 & 2.343 & -0.744 & -0.750 & 0.004 & 0.000 & 0.002 & 0.000 \\
&& 0 & 10.004 & 10.000 & 2.249 & 2.286 & -0.745 & -0.750 & 0.003 & 0.000 & 0.001 & 0.000 \\
\midrule
15 & -0.50 & 1 & 10.000 & 10.000 & 1.465 & 1.466 & -0.500 & -0.500 & 0.000 & 0.000 & 0.000 & 0.000 \\
&& 0.5 & 10.000 & 10.000 & 1.400 & 1.400 & -0.500 & -0.500 & 0.000 & 0.000 & 0.000 & 0.000 \\
&& 0.25 & 10.000 & 10.000 & 1.366 & 1.367 & -0.500 & -0.500 & 0.000 & 0.000 & 0.000 & 0.000 \\
&& 0 & 10.000 & 10.000 & 1.333 & 1.333 & -0.500 & -0.500 & 0.000 & 0.000 & 0.000 & 0.000 \\
\midrule
12.5 & -0.25 & 1 & 10.000 & 10.000 & 1.173 & 1.173 & -0.250 & -0.250 & 0.000 & 0.000 & 0.000 & 0.000 \\
&& 0.5 & 10.000 & 10.000 & 1.120 & 1.120 & -0.250 & -0.250 & 0.000 & 0.000 & 0.000 & 0.000 \\
&& 0.25 & 10.000 & 10.000 & 1.093 & 1.093 & -0.250 & -0.250 & 0.000 & 0.000 & 0.000 & 0.000 \\
&& 0 & 10.000 & 10.000 & 1.067 & 1.067 & -0.250 & -0.250 & 0.000 & 0.000 & 0.000 & 0.000 \\
\midrule
7.5 & 0.25 & 1 & 10.000 & 10.000 & 1.173 & 1.173 & 0.250 & 0.250 & 0.000 & 0.000 & 0.000 & 0.000 \\
&& 0.5 & 10.000 & 10.000 & 1.120 & 1.120 & 0.250 & 0.250 & 0.000 & 0.000 & 0.000 & 0.000 \\
&& 0.25 & 10.000 & 10.000 & 1.093 & 1.093 & 0.250 & 0.250 & 0.000 & 0.000 & 0.000 & 0.000 \\
&& 0 & 10.000 & 10.000 & 1.067 & 1.067 & 0.250 & 0.250 & 0.000 & 0.000 & 0.000 & 0.000 \\
\midrule
5 & 0.50 & 1 & 10.000 & 10.000 & 1.466 & 1.466 & 0.500 & 0.500 & 0.000 & 0.000 & 0.000 & 0.000 \\
&& 0.5 & 10.000 & 10.000 & 1.400 & 1.400 & 0.500 & 0.500 & 0.000 & 0.000 & 0.000 & 0.000 \\
&& 0.25 & 10.000 & 10.000 & 1.367 & 1.367 & 0.500 & 0.500 & 0.000 & 0.000 & 0.000 & 0.000 \\
&& 0 & 10.000 & 10.000 & 1.333 & 1.333 & 0.500 & 0.500 & 0.000 & 0.000 & 0.000 & 0.000 \\
\midrule
2.5 & 0.75 & 1 & 10.006 & 10.000 & 2.506 & 2.514 & 0.750 & 0.750 & 0.000 & 0.000 & 0.000 & 0.000 \\
&& 0.5 & 10.002 & 10.000 & 2.397 & 2.400 & 0.750 & 0.750 & 0.000 & 0.000 & 0.000 & 0.000 \\
&& 0.25 & 10.001 & 10.000 & 2.342 & 2.343 & 0.750 & 0.750 & 0.000 & 0.000 & 0.000 & 0.000 \\
&& 0 & 10.000 & 10.000 & 2.286 & 2.286 & 0.750 & 0.750 & 0.000 & 0.000 & 0.000 & 0.000 \\
\bottomrule
\end{tabular}}
\end{table}

\clearpage

\begin{table}[h]
\caption{Moment properties of STINGARCH$(1,1)$ model, see Section~\ref{On Moments and Approximate Linearity}: true values (``tob'') vs.\ linear approximation (``lin''), mean close to~5.}
\label{tabMom115}

\smallskip
\resizebox{\linewidth}{!}{
\begin{tabular}{lrrl|cc|cc|cc|cc|cc}
\toprule
 &&&& \multicolumn{2}{c|}{$\mu$} & \multicolumn{2}{c|}{$\sigma^2/\mu$} & \multicolumn{2}{c|}{ACF$(1)$} & \multicolumn{2}{c|}{ACF$(2)$} & \multicolumn{2}{c}{ACF$(3)$} \\
$\alpha_0$ & $\alpha_1$ & $\beta_1$ & $\delta$ & tob & lin & tob & lin & tob & lin & tob & lin & tob & lin \\
\midrule
8.5 & -0.45 & -0.25 & 1 & 5.013 & 5.000 & 1.623 & 1.656 & -0.512 & -0.521 & 0.356 & 0.365 & -0.247 & -0.255 \\
&&& 0.5 & 5.006 & 5.000 & 1.511 & 1.529 & -0.516 & -0.521 & 0.360 & 0.365 & -0.251 & -0.255 \\
&&& 0.25 & 5.003 & 5.000 & 1.456 & 1.464 & -0.518 & -0.521 & 0.363 & 0.365 & -0.254 & -0.255 \\
&&& 0 & 5.000 & 5.000 & 1.394 & 1.397 & -0.519 & -0.521 & 0.363 & 0.365 & -0.253 & -0.255 \\
\midrule
6 & -0.45 & 0.25 & 1 & 5.006 & 5.000 & 1.428 & 1.436 & -0.404 & -0.406 & 0.080 & 0.081 & -0.015 & -0.016 \\
&&& 0.5 & 5.004 & 5.000 & 1.322 & 1.325 & -0.405 & -0.406 & 0.080 & 0.081 & -0.015 & -0.016 \\
&&& 0.25 & 5.001 & 5.000 & 1.262 & 1.269 & -0.405 & -0.406 & 0.081 & 0.081 & -0.016 & -0.016 \\
&&& 0 & 5.000 & 5.000 & 1.212 & 1.211 & -0.407 & -0.406 & 0.080 & 0.081 & -0.014 & -0.016 \\
\midrule
4 & 0.45 & -0.25 & 1 & 5.008 & 5.000 & 1.427 & 1.436 & 0.405 & 0.406 & 0.080 & 0.081 & 0.015 & 0.016 \\
&&& 0.5 & 5.003 & 5.000 & 1.322 & 1.325 & 0.407 & 0.406 & 0.082 & 0.081 & 0.016 & 0.016 \\
&&& 0.25 & 4.995 & 5.000 & 1.266 & 1.269 & 0.406 & 0.406 & 0.081 & 0.081 & 0.016 & 0.016 \\
&&& 0 & 4.993 & 5.000 & 1.213 & 1.211 & 0.407 & 0.406 & 0.082 & 0.081 & 0.017 & 0.016 \\
\midrule
1.5 & 0.45 & 0.25 & 1 & 5.025 & 5.000 & 1.641 & 1.656 & 0.517 & 0.521 & 0.360 & 0.365 & 0.251 & 0.255 \\
&&& 0.5 & 4.999 & 5.000 & 1.522 & 1.529 & 0.520 & 0.521 & 0.364 & 0.365 & 0.255 & 0.255 \\
&&& 0.25 & 4.996 & 5.000 & 1.462 & 1.464 & 0.521 & 0.521 & 0.363 & 0.365 & 0.256 & 0.255 \\
&&& 0 & 5.003 & 5.000 & 1.396 & 1.397 & 0.520 & 0.521 & 0.364 & 0.365 & 0.254 & 0.255 \\
\midrule
8.5 & -0.25 & -0.45 & 1 & 5.006 & 5.000 & 1.324 & 1.331 & -0.296 & -0.299 & 0.206 & 0.209 & -0.144 & -0.147 \\
&&& 0.5 & 5.004 & 5.000 & 1.228 & 1.228 & -0.300 & -0.299 & 0.211 & 0.209 & -0.147 & -0.147 \\
&&& 0.25 & 5.001 & 5.000 & 1.173 & 1.176 & -0.298 & -0.299 & 0.208 & 0.209 & -0.146 & -0.147 \\
&&& 0 & 4.999 & 5.000 & 1.124 & 1.123 & -0.298 & -0.299 & 0.210 & 0.209 & -0.146 & -0.147 \\
\midrule
4 & -0.25 & 0.45 & 1 & 5.005 & 5.000 & 1.261 & 1.263 & -0.223 & -0.222 & -0.043 & -0.044 & -0.010 & -0.009 \\
&&& 0.5 & 5.001 & 5.000 & 1.168 & 1.166 & -0.222 & -0.222 & -0.046 & -0.044 & -0.008 & -0.009 \\
&&& 0.25 & 4.999 & 5.000 & 1.112 & 1.116 & -0.221 & -0.222 & -0.045 & -0.044 & -0.010 & -0.009 \\
&&& 0 & 4.999 & 5.000 & 1.064 & 1.065 & -0.222 & -0.222 & -0.044 & -0.044 & -0.008 & -0.009 \\
\midrule
6 & 0.25 & -0.45 & 1 & 5.005 & 5.000 & 1.262 & 1.263 & 0.220 & 0.222 & -0.046 & -0.044 & 0.009 & 0.009 \\
&&& 0.5 & 5.008 & 5.000 & 1.165 & 1.166 & 0.222 & 0.222 & -0.045 & -0.044 & 0.009 & 0.009 \\
&&& 0.25 & 5.006 & 5.000 & 1.117 & 1.116 & 0.221 & 0.222 & -0.044 & -0.044 & 0.011 & 0.009 \\
&&& 0 & 4.996 & 5.000 & 1.064 & 1.065 & 0.223 & 0.222 & -0.043 & -0.044 & 0.012 & 0.009 \\
\midrule
1.5 & 0.25 & 0.45 & 1 & 5.014 & 5.000 & 1.330 & 1.331 & 0.299 & 0.299 & 0.210 & 0.209 & 0.147 & 0.147 \\
&&& 0.5 & 5.008 & 5.000 & 1.226 & 1.228 & 0.298 & 0.299 & 0.208 & 0.209 & 0.146 & 0.147 \\
&&& 0.25 & 5.011 & 5.000 & 1.175 & 1.176 & 0.298 & 0.299 & 0.208 & 0.209 & 0.147 & 0.147 \\
&&& 0 & 4.996 & 5.000 & 1.125 & 1.123 & 0.300 & 0.299 & 0.210 & 0.209 & 0.146 & 0.147 \\
\bottomrule
\end{tabular}}
\end{table}

\clearpage

\begin{table}[h]
\caption{Moment properties of STINGARCH$(1,1)$ model, see Section~\ref{On Moments and Approximate Linearity}: true values (``tob'') vs.\ linear approximation (``lin''), mean close to~10.}
\label{tabMom1110}

\smallskip
\resizebox{\linewidth}{!}{
\begin{tabular}{lrrl|cc|cc|cc|cc|cc}
\toprule
 &&&& \multicolumn{2}{c|}{$\mu$} & \multicolumn{2}{c|}{$\sigma^2/\mu$} & \multicolumn{2}{c|}{ACF$(1)$} & \multicolumn{2}{c|}{ACF$(2)$} & \multicolumn{2}{c}{ACF$(3)$} \\
$\alpha_0$ & $\alpha_1$ & $\beta_1$ & $\delta$ & tob & lin & tob & lin & tob & lin & tob & lin & tob & lin \\
\midrule
17 & -0.45 & -0.25 & 1 & 10.00 & 10.00 & 1.535 & 1.537 & -0.520 & -0.521 & 0.364 & 0.365 & -0.255 & -0.255 \\
&&& 0.5 & 9.997 & 10.00 & 1.460 & 1.467 & -0.518 & -0.521 & 0.362 & 0.365 & -0.253 & -0.255 \\
&&& 0.25 & 9.998 & 10.00 & 1.430 & 1.432 & -0.520 & -0.521 & 0.364 & 0.365 & -0.255 & -0.255 \\
&&& 0 & 9.996 & 10.00 & 1.388 & 1.397 & -0.519 & -0.521 & 0.364 & 0.365 & -0.254 & -0.255 \\
\midrule
12 & -0.45 & 0.25 & 1 & 9.997 & 10.00 & 1.333 & 1.332 & -0.406 & -0.406 & 0.080 & 0.081 & -0.016 & -0.016 \\
&&& 0.5 & 9.999 & 10.00 & 1.270 & 1.271 & -0.405 & -0.406 & 0.081 & 0.081 & -0.016 & -0.016 \\
&&& 0.25 & 10.00 & 10.00 & 1.245 & 1.241 & -0.408 & -0.406 & 0.083 & 0.081 & -0.017 & -0.016 \\
&&& 0 & 10.00 & 10.00 & 1.209 & 1.211 & -0.406 & -0.406 & 0.081 & 0.081 & -0.017 & -0.016 \\
\midrule
8 & 0.45 & -0.25 & 1 & 9.991 & 10.00 & 1.331 & 1.332 & 0.406 & 0.406 & 0.080 & 0.081 & 0.018 & 0.016 \\
&&& 0.5 & 9.989 & 10.00 & 1.270 & 1.271 & 0.407 & 0.406 & 0.084 & 0.081 & 0.018 & 0.016 \\
&&& 0.25 & 10.00 & 10.00 & 1.243 & 1.241 & 0.406 & 0.406 & 0.081 & 0.081 & 0.016 & 0.016 \\
&&& 0 & 10.00 & 10.00 & 1.213 & 1.211 & 0.407 & 0.406 & 0.081 & 0.081 & 0.015 & 0.016 \\
\midrule
3 & 0.45 & 0.25 & 1 & 10.01 & 10.00 & 1.540 & 1.537 & 0.521 & 0.521 & 0.366 & 0.365 & 0.256 & 0.255 \\
&&& 0.5 & 10.01 & 10.00 & 1.464 & 1.467 & 0.520 & 0.521 & 0.365 & 0.365 & 0.256 & 0.255 \\
&&& 0.25 & 9.987 & 10.00 & 1.437 & 1.432 & 0.522 & 0.521 & 0.365 & 0.365 & 0.255 & 0.255 \\
&&& 0 & 10.01 & 10.00 & 1.402 & 1.397 & 0.521 & 0.521 & 0.365 & 0.365 & 0.255 & 0.255 \\
\midrule
17 & -0.25 & -0.45 & 1 & 9.999 & 10.00 & 1.231 & 1.235 & -0.298 & -0.299 & 0.208 & 0.209 & -0.146 & -0.147 \\
&&& 0.5 & 9.999 & 10.00 & 1.178 & 1.179 & -0.299 & -0.299 & 0.211 & 0.209 & -0.147 & -0.147 \\
&&& 0.25 & 9.996 & 10.00 & 1.150 & 1.151 & -0.299 & -0.299 & 0.210 & 0.209 & -0.147 & -0.147 \\
&&& 0 & 10.00 & 10.00 & 1.124 & 1.123 & -0.302 & -0.299 & 0.210 & 0.209 & -0.148 & -0.147 \\
\midrule
8 & -0.25 & 0.45 & 1 & 9.999 & 10.00 & 1.171 & 1.171 & -0.221 & -0.222 & -0.045 & -0.044 & -0.009 & -0.009 \\
&&& 0.5 & 10.00 & 10.00 & 1.119 & 1.118 & -0.222 & -0.222 & -0.046 & -0.044 & -0.008 & -0.009 \\
&&& 0.25 & 9.997 & 10.00 & 1.092 & 1.092 & -0.221 & -0.222 & -0.047 & -0.044 & -0.008 & -0.009 \\
&&& 0 & 9.998 & 10.00 & 1.065 & 1.065 & -0.222 & -0.222 & -0.046 & -0.044 & -0.010 & -0.009 \\
\midrule
12 & 0.25 & -0.45 & 1 & 10.00 & 10.00 & 1.172 & 1.171 & 0.221 & 0.222 & -0.045 & -0.044 & 0.009 & 0.009 \\
&&& 0.5 & 10.01 & 10.00 & 1.115 & 1.118 & 0.221 & 0.222 & -0.045 & -0.044 & 0.010 & 0.009 \\
&&& 0.25 & 9.996 & 10.00 & 1.092 & 1.092 & 0.222 & 0.222 & -0.045 & -0.044 & 0.010 & 0.009 \\
&&& 0 & 10.00 & 10.00 & 1.066 & 1.065 & 0.222 & 0.222 & -0.046 & -0.044 & 0.010 & 0.009 \\
\midrule
3 & 0.25 & 0.45 & 1 & 9.997 & 10.00 & 1.236 & 1.235 & 0.300 & 0.299 & 0.208 & 0.209 & 0.146 & 0.147 \\
&&& 0.5 & 10.01 & 10.00 & 1.179 & 1.179 & 0.299 & 0.299 & 0.210 & 0.209 & 0.147 & 0.147 \\
&&& 0.25 & 10.00 & 10.00 & 1.150 & 1.151 & 0.300 & 0.299 & 0.211 & 0.209 & 0.147 & 0.147 \\
&&& 0 & 10.00 & 10.00 & 1.122 & 1.123 & 0.300 & 0.299 & 0.210 & 0.209 & 0.147 & 0.147 \\
\bottomrule
\end{tabular}}
\end{table}

\clearpage

\begin{table}[ht]
\caption{Estimation of STINARCH$(1)$ model, see Section~\ref{Results from Simulation Study}: mean and standard error (SE) of simulated estimates, where MLE use $\delta:=0.25$ throughout.}
\label{tabEst1}

\smallskip
\resizebox{\linewidth}{!}{
\begin{tabular}{r|cc|cc|cc|cc|cc|cc}
\toprule
& \multicolumn{6}{l|}{Mean:} & \multicolumn{6}{l}{Standard error:}\\
& \multicolumn{2}{c|}{MLE} & \multicolumn{2}{c|}{CLADE} & \multicolumn{2}{c|}{CLS} & \multicolumn{2}{c|}{MLE} & \multicolumn{2}{c|}{CLADE} & \multicolumn{2}{c}{CLS} \\
$n$ & $\hat{\alpha}_0$ & $\hat{\alpha}_1$ & $\check{\alpha}_0$ & $\check{\alpha}_1$ & $\breve{\alpha}_0$ & $\breve{\alpha}_1$ & $\hat{\alpha}_0$ & $\hat{\alpha}_1$ & $\check{\alpha}_0$ & $\check{\alpha}_1$ & $\breve{\alpha}_0$ & $\breve{\alpha}_1$ \\
\midrule
\multicolumn{13}{l}{$(\alpha_0,\alpha_1)=(7.5,-0.5)$ with true $\delta=0$} \\
\midrule
100 & 7.470 & -0.497 & 7.336 & -0.500 & 7.468 & -0.496 & 0.498 & 0.077 & 0.488 & 0.073 & 0.518 & 0.083 \\
250 & 7.505 & -0.503 & 7.354 & -0.499 & 7.497 & -0.500 & 0.303 & 0.046 & 0.294 & 0.037 & 0.326 & 0.052 \\
500 & 7.511 & -0.503 & 7.392 & -0.500 & 7.493 & -0.498 & 0.217 & 0.033 & 0.228 & 0.023 & 0.234 & 0.037 \\
1000 & 7.510 & -0.503 & 7.436 & -0.500 & 7.495 & -0.499 & 0.149 & 0.022 & 0.172 & 0.013 & 0.162 & 0.026 \\
\midrule
\multicolumn{13}{l}{$(\alpha_0,\alpha_1)=(7.5,-0.5)$ with true $\delta=0.25$} \\
\midrule
100 & 7.483 & -0.496 & 7.343 & -0.497 & 7.474 & -0.494 & 0.487 & 0.077 & 0.513 & 0.076 & 0.517 & 0.084 \\
250 & 7.496 & -0.500 & 7.372 & -0.499 & 7.490 & -0.498 & 0.301 & 0.047 & 0.282 & 0.037 & 0.320 & 0.052 \\
500 & 7.502 & -0.500 & 7.394 & -0.500 & 7.493 & -0.498 & 0.220 & 0.032 & 0.227 & 0.023 & 0.237 & 0.037 \\
1000 & 7.499 & -0.500 & 7.427 & -0.499 & 7.493 & -0.498 & 0.156 & 0.023 & 0.183 & 0.015 & 0.169 & 0.027 \\
\midrule
\multicolumn{13}{l}{$(\alpha_0,\alpha_1)=(7.5,-0.5)$ with true $\delta=1$} \\
\midrule
100 & 7.469 & -0.491 & 7.360 & -0.498 & 7.482 & -0.494 & 0.523 & 0.077 & 0.563 & 0.085 & 0.559 & 0.088 \\
250 & 7.455 & -0.489 & 7.389 & -0.499 & 7.465 & -0.492 & 0.317 & 0.048 & 0.302 & 0.040 & 0.343 & 0.055 \\
500 & 7.472 & -0.491 & 7.406 & -0.499 & 7.480 & -0.493 & 0.226 & 0.034 & 0.216 & 0.024 & 0.243 & 0.038 \\
1000 & 7.463 & -0.490 & 7.451 & -0.500 & 7.479 & -0.494 & 0.152 & 0.023 & 0.155 & 0.014 & 0.163 & 0.025 \\
\midrule
\multicolumn{13}{l}{$(\alpha_0,\alpha_1)=(2.5,0.5)$ with true $\delta=0$} \\
\midrule
100 & 2.601 & 0.478 & 2.460 & 0.477 & 2.627 & 0.473 & 0.479 & 0.095 & 0.459 & 0.091 & 0.485 & 0.097 \\
250 & 2.539 & 0.490 & 2.433 & 0.484 & 2.555 & 0.488 & 0.293 & 0.058 & 0.288 & 0.056 & 0.302 & 0.060 \\
500 & 2.509 & 0.498 & 2.450 & 0.491 & 2.524 & 0.495 & 0.194 & 0.039 & 0.182 & 0.034 & 0.201 & 0.040 \\
1000 & 2.500 & 0.499 & 2.467 & 0.494 & 2.511 & 0.498 & 0.145 & 0.029 & 0.135 & 0.022 & 0.150 & 0.030 \\
\midrule
\multicolumn{13}{l}{$(\alpha_0,\alpha_1)=(2.5,0.5)$ with true $\delta=0.25$} \\
\midrule
100 & 2.613 & 0.474 & 2.467 & 0.474 & 2.636 & 0.470 & 0.476 & 0.093 & 0.467 & 0.096 & 0.480 & 0.095 \\
250 & 2.546 & 0.490 & 2.444 & 0.487 & 2.551 & 0.490 & 0.292 & 0.057 & 0.260 & 0.051 & 0.306 & 0.061 \\
500 & 2.523 & 0.495 & 2.454 & 0.490 & 2.533 & 0.493 & 0.205 & 0.040 & 0.178 & 0.034 & 0.211 & 0.042 \\
1000 & 2.517 & 0.497 & 2.471 & 0.494 & 2.523 & 0.496 & 0.145 & 0.028 & 0.139 & 0.023 & 0.152 & 0.030 \\
\midrule
\multicolumn{13}{l}{$(\alpha_0,\alpha_1)=(2.5,0.5)$ with true $\delta=1$} \\
\midrule
100 & 2.644 & 0.471 & 2.499 & 0.472 & 2.660 & 0.467 & 0.483 & 0.094 & 0.489 & 0.098 & 0.495 & 0.097 \\
250 & 2.568 & 0.489 & 2.454 & 0.487 & 2.573 & 0.488 & 0.298 & 0.057 & 0.294 & 0.055 & 0.311 & 0.060 \\
500 & 2.549 & 0.493 & 2.465 & 0.491 & 2.550 & 0.492 & 0.205 & 0.041 & 0.184 & 0.036 & 0.215 & 0.043 \\
1000 & 2.546 & 0.494 & 2.475 & 0.494 & 2.544 & 0.494 & 0.144 & 0.028 & 0.129 & 0.023 & 0.149 & 0.030 \\
\bottomrule
\end{tabular}}
\end{table}

\clearpage

\begin{table}[ht]
\caption{Estimation of STINARCH$(1)$ model, see Section~\ref{Results from Simulation Study}: standard error (SE) of simulated MLE and mean of approximate SE, where MLE use $\delta:=0.25$ throughout.}
\label{tabEst1b}

\smallskip
\resizebox{\linewidth}{!}{
\begin{tabular}{r|cc|cc|cc|cc|cc|cc}
\toprule
& \multicolumn{4}{l|}{True $\delta=0$:} & \multicolumn{4}{l|}{True $\delta=0.25$:} & \multicolumn{4}{l}{True $\delta=1$:}\\
& \multicolumn{2}{c|}{simul.\ SE} & \multicolumn{2}{c|}{approx.\ SE} & \multicolumn{2}{c|}{simul.\ SE} & \multicolumn{2}{c|}{approx.\ SE} & \multicolumn{2}{c|}{simul.\ SE} & \multicolumn{2}{c}{approx.\ SE} \\
$n$ & $\hat{\alpha}_0$ & $\hat{\alpha}_1$ & $\hat{\alpha}_0$ & $\hat{\alpha}_1$ & $\hat{\alpha}_0$ & $\hat{\alpha}_1$ & $\hat{\alpha}_0$ & $\hat{\alpha}_1$ & $\hat{\alpha}_0$ & $\hat{\alpha}_1$ & $\hat{\alpha}_0$ & $\hat{\alpha}_1$ \\
\midrule
\multicolumn{13}{l}{$(\alpha_0,\alpha_1)=(7.5,-0.5)$} \\
\midrule
100 & 0.498 & 0.077 & 0.499 & 0.078 & 0.487 & 0.077 & 0.522 & 0.081 & 0.523 & 0.077 & 0.530 & 0.082 \\
250 & 0.303 & 0.046 & 0.306 & 0.047 & 0.301 & 0.047 & 0.316 & 0.048 & 0.317 & 0.048 & 0.324 & 0.048 \\
500 & 0.217 & 0.033 & 0.215 & 0.032 & 0.220 & 0.032 & 0.211 & 0.032 & 0.226 & 0.034 & 0.220 & 0.032 \\
1000 & 0.149 & 0.022 & 0.147 & 0.022 & 0.156 & 0.023 & 0.155 & 0.023 & 0.152 & 0.023 & 0.156 & 0.023 \\
\midrule
\multicolumn{13}{l}{$(\alpha_0,\alpha_1)=(2.5,0.5)$} \\
\midrule
100 & 0.479 & 0.095 & 0.473 & 0.094 & 0.476 & 0.093 & 0.474 & 0.095 & 0.483 & 0.094 & 0.488 & 0.094 \\
250 & 0.293 & 0.058 & 0.288 & 0.058 & 0.292 & 0.057 & 0.284 & 0.056 & 0.298 & 0.057 & 0.289 & 0.057 \\
500 & 0.194 & 0.039 & 0.202 & 0.041 & 0.205 & 0.040 & 0.198 & 0.039 & 0.205 & 0.041 & 0.207 & 0.041 \\
1000 & 0.145 & 0.029 & 0.141 & 0.027 & 0.145 & 0.028 & 0.140 & 0.028 & 0.144 & 0.028 & 0.141 & 0.028 \\
\bottomrule
\end{tabular}}
\end{table}


\begin{table}[h!]
\caption{Estimation of STINARCH$(1)$ model, see Section~\ref{Results from Simulation Study}, where also~$\delta$ is estimated: mean and standard error (SE) of simulated MLE as well as mean of approximate SE.}
\label{tabEst1delta}

\smallskip
\resizebox{\linewidth}{!}{
\begin{tabular}{lrlr|crc|crc|crc}
\toprule
&&&& \multicolumn{3}{l|}{Mean:} & \multicolumn{3}{l|}{Standard error:} & \multicolumn{3}{l}{Mean approx.\ SE:}\\
$\alpha_0$ & $\alpha_1$ & $\delta$ & $n$ & $\hat{\alpha}_0$ & $\hat{\alpha}_1$ & $\hat{\delta}$ & $\hat{\alpha}_0$ & $\hat{\alpha}_1$ & $\hat{\delta}$ & $\hat{\alpha}_0$ & $\hat{\alpha}_1$ & $\hat{\delta}$ \\
\midrule
7.5 & -0.5 & 0.25 & 100 & 7.495 & -0.499 & 0.343 & 0.498 & 0.076 & 0.449 & 0.497 & 0.077 & 0.445 \\
 &  &  & 250 & 7.499 & -0.499 & 0.273 & 0.302 & 0.046 & 0.295 & 0.295 & 0.045 & 0.297 \\
 &  &  & 500 & 7.498 & -0.500 & 0.258 & 0.229 & 0.034 & 0.226 & 0.223 & 0.034 & 0.202 \\
 &  &  & 1000 & 7.508 & -0.501 & 0.254 & 0.153 & 0.023 & 0.172 & 0.156 & 0.023 & 0.164 \\
\midrule
7.5 & -0.5 & 1 & 100 & 7.495 & -0.496 & 0.945 & 0.517 & 0.080 & 0.745 & 0.534 & 0.083 & 0.677 \\
 &  &  & 250 & 7.477 & -0.496 & 0.947 & 0.329 & 0.049 & 0.469 & 0.323 & 0.050 & 0.460 \\
 &  &  & 500 & 7.501 & -0.501 & 0.982 & 0.224 & 0.035 & 0.335 & 0.224 & 0.034 & 0.340 \\
 &  &  & 1000 & 7.494 & -0.499 & 0.987 & 0.162 & 0.024 & 0.242 & 0.159 & 0.025 & 0.251 \\
\midrule
2.5 & 0.5 & 0.25 & 100 & 2.630 & 0.473 & 0.312 & 0.474 & 0.091 & 0.426 & 0.446 & 0.090 & 0.431 \\
 &  &  & 250 & 2.554 & 0.488 & 0.284 & 0.288 & 0.058 & 0.286 & 0.282 & 0.057 & 0.283 \\
 &  &  & 500 & 2.526 & 0.493 & 0.264 & 0.203 & 0.040 & 0.221 & 0.199 & 0.039 & 0.217 \\
 &  &  & 1000 & 2.509 & 0.498 & 0.261 & 0.141 & 0.028 & 0.166 & 0.136 & 0.028 & 0.159 \\
\midrule
2.5 & 0.5 & 1 & 100 & 2.614 & 0.475 & 0.960 & 0.473 & 0.089 & 0.736 & 0.480 & 0.093 & 0.690 \\
 &  &  & 250 & 2.530 & 0.492 & 0.967 & 0.312 & 0.060 & 0.486 & 0.298 & 0.057 & 0.459 \\
 &  &  & 500 & 2.521 & 0.496 & 0.982 & 0.209 & 0.042 & 0.329 & 0.207 & 0.041 & 0.333 \\
 &  &  & 1000 & 2.509 & 0.497 & 0.995 & 0.144 & 0.029 & 0.240 & 0.147 & 0.028 & 0.241 \\
\bottomrule
\end{tabular}}
\end{table}

\clearpage

\begin{table}[ht]
\caption{Estimation of STINGARCH$(1,1)$ model, see Section~\ref{Results from Simulation Study}: mean and standard error (SE) of simulated estimates, where MLE use $\delta:=0.25$ throughout.}
\label{tabEst11}

\smallskip
\resizebox{\linewidth}{!}{
\begin{tabular}{r|ccc|ccc|ccc|ccc}
\toprule
& \multicolumn{6}{l|}{Mean:} & \multicolumn{6}{l}{Standard error:}\\
& \multicolumn{3}{c|}{MLE} & \multicolumn{3}{c|}{CLADE} & \multicolumn{3}{c|}{MLE} & \multicolumn{3}{c}{CLADE} \\
$n$ & $\hat{\alpha}_0$ & $\hat{\alpha}_1$ & $\hat{\beta}_1$ & $\check{\alpha}_0$ & $\check{\alpha}_1$ & $\check{\beta}_1$ & $\hat{\alpha}_0$ & $\hat{\alpha}_1$ & $\hat{\beta}_1$ & $\check{\alpha}_0$ & $\check{\alpha}_1$ & $\check{\beta}_1$ \\
\midrule
\multicolumn{13}{l}{$(\alpha_0,\alpha_1,\beta_1)=(7.25,-0.45,0.25)$ with true $\delta=0$} \\
\midrule
250 & 7.152 & -0.451 & 0.267 & 7.041 & -0.450 & 0.264 & 0.937 & 0.059 & 0.153 & 1.032 & 0.076 & 0.173 \\
500 & 7.192 & -0.451 & 0.259 & 7.091 & -0.451 & 0.256 & 0.638 & 0.040 & 0.105 & 0.788 & 0.051 & 0.134 \\
1000 & 7.214 & -0.452 & 0.258 & 7.104 & -0.451 & 0.255 & 0.431 & 0.031 & 0.071 & 0.565 & 0.038 & 0.091 \\
\midrule
\multicolumn{13}{l}{$(\alpha_0,\alpha_1,\beta_1)=(7.25,-0.45,0.25)$ with true $\delta=0.25$} \\
\midrule
250 & 7.129 & -0.452 & 0.271 & 7.048 & -0.453 & 0.267 & 0.911 & 0.060 & 0.153 & 0.976 & 0.076 & 0.166 \\
500 & 7.175 & -0.453 & 0.266 & 7.098 & -0.453 & 0.259 & 0.610 & 0.041 & 0.100 & 0.768 & 0.052 & 0.128 \\
1000 & 7.243 & -0.452 & 0.253 & 7.136 & -0.452 & 0.251 & 0.439 & 0.028 & 0.073 & 0.563 & 0.036 & 0.093 \\
\midrule
\multicolumn{13}{l}{$(\alpha_0,\alpha_1,\beta_1)=(7.25,-0.45,0.25)$ with true $\delta=1$} \\
\midrule
250 & 7.113 & -0.449 & 0.272 & 7.064 & -0.451 & 0.264 & 0.930 & 0.060 & 0.154 & 1.000 & 0.074 & 0.170 \\
500 & 7.158 & -0.451 & 0.266 & 7.100 & -0.452 & 0.259 & 0.625 & 0.041 & 0.102 & 0.743 & 0.054 & 0.125 \\
1000 & 7.253 & -0.448 & 0.248 & 7.170 & -0.450 & 0.245 & 0.456 & 0.028 & 0.076 & 0.573 & 0.037 & 0.097 \\
\midrule
\multicolumn{13}{l}{$(\alpha_0,\alpha_1,\beta_1)=(2.75,0.45,-0.25)$ with true $\delta=0$} \\
\midrule
250 & 2.764 & 0.450 & -0.257 & 2.549 & 0.446 & -0.248 & 0.531 & 0.068 & 0.151 & 0.637 & 0.089 & 0.183 \\
500 & 2.758 & 0.451 & -0.258 & 2.510 & 0.448 & -0.241 & 0.368 & 0.045 & 0.101 & 0.514 & 0.056 & 0.145 \\
1000 & 2.762 & 0.450 & -0.256 & 2.532 & 0.449 & -0.248 & 0.246 & 0.033 & 0.069 & 0.372 & 0.044 & 0.105 \\
\midrule
\multicolumn{13}{l}{$(\alpha_0,\alpha_1,\beta_1)=(2.75,0.45,-0.25)$ with true $\delta=0.25$} \\
\midrule
250 & 2.801 & 0.448 & -0.264 & 2.598 & 0.445 & -0.257 & 0.525 & 0.066 & 0.146 & 0.626 & 0.080 & 0.178 \\
500 & 2.806 & 0.445 & -0.262 & 2.568 & 0.444 & -0.250 & 0.371 & 0.046 & 0.100 & 0.491 & 0.059 & 0.135 \\
1000 & 2.765 & 0.448 & -0.252 & 2.537 & 0.446 & -0.243 & 0.264 & 0.031 & 0.072 & 0.388 & 0.040 & 0.107 \\
\midrule
\multicolumn{13}{l}{$(\alpha_0,\alpha_1,\beta_1)=(2.75,0.45,-0.25)$ with true $\delta=1$} \\
\midrule
250 & 2.854 & 0.443 & -0.265 & 2.639 & 0.443 & -0.258 & 0.538 & 0.065 & 0.148 & 0.656 & 0.080 & 0.186 \\
500 & 2.840 & 0.438 & -0.257 & 2.599 & 0.441 & -0.248 & 0.378 & 0.046 & 0.103 & 0.478 & 0.060 & 0.137 \\
1000 & 2.803 & 0.442 & -0.248 & 2.577 & 0.442 & -0.241 & 0.271 & 0.031 & 0.072 & 0.339 & 0.039 & 0.098 \\
\bottomrule
\end{tabular}}
\end{table}

\clearpage

\begin{table}[ht]
\caption{Estimation of STINGARCH$(1,1)$ model, see Section~\ref{Results from Simulation Study}: mean and standard error (SE) of simulated estimates, where MLE use $\delta:=0.25$ throughout.}
\label{tabEst112}

\smallskip
\resizebox{\linewidth}{!}{
\begin{tabular}{r|ccc|ccc|ccc|ccc}
\toprule
& \multicolumn{6}{l|}{Means:} & \multicolumn{6}{l}{Standard errors:}\\
& \multicolumn{3}{c|}{MLE} & \multicolumn{3}{c|}{CLADE} & \multicolumn{3}{c|}{MLE} & \multicolumn{3}{c}{CLADE} \\
$n$ & $\hat{\alpha}_0$ & $\hat{\alpha}_1$ & $\hat{\beta}_1$ & $\check{\alpha}_0$ & $\check{\alpha}_1$ & $\check{\beta}_1$ & $\hat{\alpha}_0$ & $\hat{\alpha}_1$ & $\hat{\beta}_1$ & $\check{\alpha}_0$ & $\check{\alpha}_1$ & $\check{\beta}_1$ \\
\midrule
\multicolumn{13}{l}{$(\alpha_0,\alpha_1,\beta_1)=(6.25,-0.25,0.45)$ with true $\delta=0$} \\
\midrule
250 & 6.460 & -0.260 & 0.434 & 6.358 & -0.260 & 0.435 & 1.506 & 0.060 & 0.185 & 1.220 & 0.075 & 0.159 \\
500 & 6.335 & -0.253 & 0.442 & 6.285 & -0.257 & 0.440 & 1.161 & 0.042 & 0.141 & 1.087 & 0.050 & 0.138 \\
1000 & 6.278 & -0.254 & 0.450 & 6.214 & -0.253 & 0.446 & 0.838 & 0.031 & 0.098 & 0.901 & 0.039 & 0.108 \\
\midrule
\multicolumn{13}{l}{$(\alpha_0,\alpha_1,\beta_1)=(6.25,-0.25,0.45)$ with true $\delta=0.25$} \\
\midrule
250 & 6.368 & -0.263 & 0.448 & 6.278 & -0.262 & 0.447 & 1.422 & 0.062 & 0.173 & 1.149 & 0.077 & 0.147 \\
500 & 6.366 & -0.256 & 0.442 & 6.306 & -0.256 & 0.437 & 1.147 & 0.044 & 0.137 & 1.084 & 0.054 & 0.135 \\
1000 & 6.310 & -0.252 & 0.444 & 6.207 & -0.254 & 0.447 & 0.848 & 0.031 & 0.102 & 0.942 & 0.038 & 0.116 \\
\midrule
\multicolumn{13}{l}{$(\alpha_0,\alpha_1,\beta_1)=(6.25,-0.25,0.45)$ with true $\delta=1$} \\
\midrule
250 & 6.417 & -0.261 & 0.440 & 6.298 & -0.258 & 0.442 & 1.451 & 0.062 & 0.181 & 1.177 & 0.078 & 0.151 \\
500 & 6.391 & -0.256 & 0.439 & 6.354 & -0.254 & 0.429 & 1.154 & 0.044 & 0.142 & 1.149 & 0.056 & 0.144 \\
1000 & 6.326 & -0.252 & 0.443 & 6.261 & -0.253 & 0.441 & 0.872 & 0.030 & 0.103 & 0.950 & 0.038 & 0.115 \\
\midrule
\multicolumn{13}{l}{$(\alpha_0,\alpha_1,\beta_1)=(3.75,0.25,-0.45)$ with true $\delta=0$} \\
\midrule
250 & 3.683 & 0.255 & -0.436 & 3.543 & 0.209 & -0.414 & 0.712 & 0.065 & 0.210 & 0.669 & 0.138 & 0.200 \\
500 & 3.725 & 0.253 & -0.451 & 3.555 & 0.209 & -0.419 & 0.516 & 0.043 & 0.151 & 0.590 & 0.115 & 0.181 \\
1000 & 3.744 & 0.251 & -0.453 & 3.608 & 0.204 & -0.427 & 0.347 & 0.032 & 0.100 & 0.442 & 0.109 & 0.141 \\
\midrule
\multicolumn{13}{l}{$(\alpha_0,\alpha_1,\beta_1)=(3.75,0.25,-0.45)$ with true $\delta=0.25$} \\
\midrule
250 & 3.664 & 0.256 & -0.428 & 3.504 & 0.217 & -0.404 & 0.708 & 0.065 & 0.209 & 0.702 & 0.131 & 0.211 \\
500 & 3.771 & 0.248 & -0.455 & 3.594 & 0.213 & -0.431 & 0.501 & 0.044 & 0.144 & 0.576 & 0.108 & 0.172 \\
1000 & 3.758 & 0.249 & -0.451 & 3.617 & 0.214 & -0.439 & 0.369 & 0.031 & 0.104 & 0.457 & 0.095 & 0.137 \\
\midrule
\multicolumn{13}{l}{$(\alpha_0,\alpha_1,\beta_1)=(3.75,0.25,-0.45)$ with true $\delta=1$} \\
\midrule
250 & 3.733 & 0.252 & -0.434 & 3.537 & 0.229 & -0.417 & 0.705 & 0.062 & 0.204 & 0.731 & 0.119 & 0.222 \\
500 & 3.821 & 0.244 & -0.454 & 3.600 & 0.221 & -0.430 & 0.531 & 0.045 & 0.151 & 0.615 & 0.102 & 0.186 \\
1000 & 3.773 & 0.246 & -0.441 & 3.576 & 0.232 & -0.433 & 0.385 & 0.033 & 0.109 & 0.496 & 0.079 & 0.144 \\
\bottomrule
\end{tabular}}
\end{table}

\clearpage

\begin{table}[ht]
\caption{Estimation of STINGARCH$(1,1)$ model, see Section~\ref{Results from Simulation Study}: standard error of simulated MLE and mean of approximate SE, where MLE use $\delta:=0.25$ throughout.}
\label{tabEst113}

\smallskip
\resizebox{\linewidth}{!}{
\begin{tabular}{r|ccc|ccc|ccc|ccc}
\toprule
& \multicolumn{3}{c|}{Simulated SE:} & \multicolumn{3}{c|}{Approximate SE:} & \multicolumn{3}{c|}{Simulated SE:} & \multicolumn{3}{c}{Approximate SE:} \\
$n$ & $\hat{\alpha}_0$ & $\hat{\alpha}_1$ & $\hat{\beta}_1$ & $\hat{\alpha}_0$ & $\hat{\alpha}_1$ & $\hat{\beta}_1$ & $\hat{\alpha}_0$ & $\hat{\alpha}_1$ & $\hat{\beta}_1$ & $\hat{\alpha}_0$ & $\hat{\alpha}_1$ & $\hat{\beta}_1$ \\
\midrule
\multicolumn{7}{l}{$(\alpha_0,\alpha_1,\beta_1)=(7.25,-0.45,0.25)$ with true $\delta=0$} & \multicolumn{6}{l}{$(6.25,-0.25,0.45)$ with true $\delta=0$} \\
\midrule
250 & 0.937 & 0.059 & 0.153 & 0.942 & 0.061 & 0.152 & 1.506 & 0.060 & 0.185 & 1.425 & 0.061 & 0.171 \\
500 & 0.638 & 0.040 & 0.105 & 0.633 & 0.039 & 0.103 & 1.161 & 0.042 & 0.141 & 1.154 & 0.042 & 0.140 \\
1000 & 0.431 & 0.031 & 0.071 & 0.438 & 0.028 & 0.074 & 0.838 & 0.031 & 0.098 & 0.849 & 0.031 & 0.103 \\
\midrule
\multicolumn{7}{l}{$(\alpha_0,\alpha_1,\beta_1)=(7.25,-0.45,0.25)$ with true $\delta=0.25$} & \multicolumn{6}{l}{$(6.25,-0.25,0.45)$ with true $\delta=0.25$} \\
\midrule
250 & 0.911 & 0.060 & 0.153 & 0.905 & 0.059 & 0.149 & 1.422 & 0.062 & 0.173 & 1.445 & 0.061 & 0.177 \\
500 & 0.610 & 0.041 & 0.100 & 0.629 & 0.040 & 0.104 & 1.147 & 0.044 & 0.137 & 1.132 & 0.043 & 0.135 \\
1000 & 0.439 & 0.028 & 0.073 & 0.446 & 0.029 & 0.073 & 0.848 & 0.031 & 0.102 & 0.880 & 0.031 & 0.104 \\
\midrule
\multicolumn{7}{l}{$(\alpha_0,\alpha_1,\beta_1)=(7.25,-0.45,0.25)$ with true $\delta=1$} & \multicolumn{6}{l}{$(6.25,-0.25,0.45)$ with true $\delta=1$} \\
\midrule
250 & 0.930 & 0.060 & 0.154 & 0.931 & 0.062 & 0.151 & 1.451 & 0.062 & 0.181 & 1.455 & 0.060 & 0.178 \\
500 & 0.625 & 0.041 & 0.102 & 0.607 & 0.042 & 0.101 & 1.154 & 0.044 & 0.142 & 1.152 & 0.043 & 0.137 \\
1000 & 0.456 & 0.028 & 0.076 & 0.435 & 0.030 & 0.072 & 0.872 & 0.030 & 0.103 & 0.869 & 0.032 & 0.102 \\
\midrule
\multicolumn{7}{l}{$(\alpha_0,\alpha_1,\beta_1)=(2.75,0.45,-0.25)$ with true $\delta=0$} & \multicolumn{6}{l}{$(3.75,0.25,-0.45)$ with true $\delta=0$} \\
\midrule
250 & 0.531 & 0.068 & 0.151 & 0.529 & 0.066 & 0.149 & 0.712 & 0.065 & 0.210 & 0.715 & 0.065 & 0.213 \\
500 & 0.368 & 0.045 & 0.101 & 0.356 & 0.045 & 0.100 & 0.516 & 0.043 & 0.151 & 0.500 & 0.044 & 0.146 \\
1000 & 0.246 & 0.033 & 0.069 & 0.256 & 0.031 & 0.070 & 0.347 & 0.032 & 0.100 & 0.368 & 0.031 & 0.105 \\
\midrule
\multicolumn{7}{l}{$(\alpha_0,\alpha_1,\beta_1)=(2.75,0.45,-0.25)$ with true $\delta=0.25$} & \multicolumn{6}{l}{$(3.75,0.25,-0.45)$ with true $\delta=0.25$} \\
\midrule
250 & 0.525 & 0.066 & 0.146 & 0.523 & 0.066 & 0.148 & 0.708 & 0.065 & 0.209 & 0.708 & 0.063 & 0.212 \\
500 & 0.371 & 0.046 & 0.100 & 0.362 & 0.047 & 0.102 & 0.501 & 0.044 & 0.144 & 0.513 & 0.046 & 0.148 \\
1000 & 0.264 & 0.031 & 0.072 & 0.253 & 0.032 & 0.069 & 0.369 & 0.031 & 0.104 & 0.373 & 0.032 & 0.107 \\
\midrule
\multicolumn{7}{l}{$(\alpha_0,\alpha_1,\beta_1)=(2.75,0.45,-0.25)$ with true $\delta=1$} & \multicolumn{6}{l}{$(3.75,0.25,-0.45)$ with true $\delta=1$} \\
\midrule
250 & 0.538 & 0.065 & 0.148 & 0.550 & 0.065 & 0.148 & 0.705 & 0.062 & 0.204 & 0.775 & 0.065 & 0.234 \\
500 & 0.378 & 0.046 & 0.103 & 0.375 & 0.045 & 0.102 & 0.531 & 0.045 & 0.151 & 0.525 & 0.046 & 0.150 \\
1000 & 0.271 & 0.031 & 0.072 & 0.261 & 0.033 & 0.070 & 0.385 & 0.033 & 0.109 & 0.374 & 0.032 & 0.106 \\
\bottomrule
\end{tabular}}
\end{table}


\begin{thebibliography}{}

\bibitem[Agosto et al.(2016)]{agosto16}
Agosto, A., Cavaliere, G., Kristensen, D., Rahbek, A. (2016)
Modeling corporate defaults: Poisson autoregressions with exogenous covariates (PARX).
\textit{Journal of Empirical Finance} \textbf{38}, 640--663.

\bibitem[Alzaid \& Al-Osh(1990)]{alzaid90}
Alzaid, A.A., Al-Osh, M.A. (1990)
An integer-valued pth-order autoregressive structure (INAR($p$)) process.
\textit{Journal of Applied Probability} \textbf{27}, 314--324.

\bibitem[Alzaid \& Omair(2014)]{alzaid14}
Alzaid, A.A., Omair, M.A. (2014)
Poisson difference integer valued autoregressive model of order one.
\textit{Bulletin of the Malaysian Mathematical Science Society} \textbf{37}, 465--485.

\bibitem[Amemiya(1985)]{amemiya85}
Amemiya, T. (1985)
\textit{Advanced Econometrics}.
Harvard University Press, Cambridge.

\bibitem[Andersson \& Karlis(2014)]{andersson14}
Andersson, J., Karlis, D. (2014)
A parametric time series model with covariates for integers in $\bbz$.
\textit{Statistical Modelling} \textbf{14}, 135--156.

\bibitem[Balachandran et al.(2017)]{balachandran17}
Balachandran, P., Kolaczyk, E.D., Viles, W.D. (2017)
On the propagation of low-rate measurement error to subgraph counts in large networks.
\textit{Journal of Machine Learning Research} \textbf{18}, 1--33.

\bibitem[Bilias et al.(2019)]{bilias19}
Bilias, Y.,  Florios, K., Skouras, S. (2019)
Exact computation of Censored Least Absolute Deviations estimator.
\textit{Journal of Econometrics} \textbf{212}, 584--606.

\bibitem[Bykhovskaya(2023)]{bykhovskaya23}
Bykhovskaya, A. (2023)
Time series approach to the evolution of networks: prediction and estimation.
\textit{Journal of Business \& Economic Statistics} \textbf{41}, 170--183.

\bibitem[Cai et al.(2017)]{cai17}
Cai, Z., He, X., Sun, J., Vasconcelos, N. (2017)
Deep learning with low precision by half-wave Gaussian quantization.
\textit{2017 IEEE Conference on Computer Vision and Pattern Recognition (CVPR)}, Honolulu, HI, 5406--5414.

\bibitem[Carallo et al.(2024)]{carallo20}
Carallo, G., Casarin, R., Robert, C.P. (2024)
Generalized Poisson difference autoregressive processes.
\textit{International Journal of Forecasting}, forthcoming, \url{https://doi.org/10.1016/j.ijforecast.2023.11.009}

\bibitem[Chen et al.(2018)]{chen18}
Chen, C.W.S., Hsieh, Y., Su, H., Wu, J.J. (2018)
Causality test of ambient fine particles and human influenza in Taiwan: Age group-specific disparity and geographic heterogeneity.
\textit{Environment International} \textbf{111}, 354--361.

\bibitem[de~Jong \& Herrera(2011)]{dejong11}
de Jong, R., Herrera, A.M. (2011)
Dynamic censored regression and the open market desk reaction function.
\textit{Journal of Business \& Economic Statistics} \textbf{29}, 228--237.

\bibitem[Doukhan et al.(2021)]{doukhan21}
Doukhan, P., Mamode Khan, N., Neumann, M.H. (2021)
Mixing properties of integer-valued GARCH processes.
\textit{ALEA} \textbf{18}, 401--420.

\bibitem[Du \& Li(1991)]{duli91}
Du, J.-G., Li, Y. (1991)
The integer-valued autoregressive (INAR($p$)) model.
\textit{Journal of Time Series Analysis} \textbf{12}, 129--142.

\bibitem[Ferland et al.(2006)]{ferland06}
Ferland, R., Latour, A., Oraichi, D. (2006)
Integer-valued GARCH processes.
\textit{Journal of Time Series Analysis} \textbf{27}, 923--942.

\bibitem[Fokianos et al.(2009)]{fokianos09}
Fokianos, K., Rahbek, A., Tj\o stheim, D. (2009)
Poisson autoregression.
\textit{Journal of the American Statistical Association} \textbf{104}, 1430--1439.

\bibitem[Fokianos \& Tj\o stheim(2011)]{fokianos11}
Fokianos, K., Tj\o stheim, D. (2011)
Log-linear Poisson autoregression.
\textit{Journal of Multivariate Analysis} \textbf{102}, 563--578.

\bibitem[Goldberger(1964)]{goldberger64}
Goldberger, A.S. (1964)
\textit{Econometric Theory}.
John Wiley \& Sons, New York.

\bibitem[Grunwald et al.(2000)]{grunwald00}
Grunwald, G., Hyndman, R.J., Tedesco, L., Tweedie, R.L. (2000)
Non-Gaussian conditional linear AR(1) models.
\textit{Australian and New Zealand Journal of Statistics} \textbf{42}, 479--495.

\bibitem[Honor\'e(1993)]{honore93}
Honor\'e, B.E. (1993)
Orthogonality conditions for Tobit models with fixed effects and lagged dependent variables.
\textit{Journal of Econometrics} \textbf{59}, 35--61.

\bibitem[Honor\'e \& Hu(2004)]{honore04}
Honor\'e, B.E., Hu, L. (2004)
Estimation of cross sectional and panel data censored regression models with endogeneity.
\textit{Journal of Econometrics} \textbf{122}, 293--316.

\bibitem[Jeffrey \& Dai(2008)]{jeffrey08}
Jeffrey, A., Dai, H. (2008)
\textit{Handbook of Mathematical Formulas and Integrals}.
4\textsuperscript{th} edition, Academic Press, San Diego.

\bibitem[Johnson(1959)]{johnson59}
Johnson, N.L. (1959)
On an extension of the connexion between Poisson and $\chi^2$ distributions.
\textit{Biometrika} \textbf{46}, 352--363.

\bibitem[Johnson et al.(2005)]{johnson05}
Johnson, N.L., Kemp, A.W., Kotz, S. (2005)
\textit{Univariate Discrete Distributions}.
3\textsuperscript{rd} edition, John Wiley \& Sons, New Jersey.

\bibitem[Karlis \& Ntzoufras(2006)]{karlis06}
Karlis, D., Ntzoufras, I. (2006)
Bayesian analysis of the differences of count data.
\textit{Statistics in Medicine} \textbf{25}, 1885--1905.

\bibitem[Karlis \& Ntzoufras(2009)]{karlis09}
Karlis, D., Ntzoufras, I. (2009)
Bayesian modelling of football outcomes: using the Skellam's distribution for the goal difference.
\textit{IMA Journal of Management Mathematics} \textbf{20}, 133--145.

\bibitem[Kim \& Park(2008)]{kim08}
Kim, H.-Y., Park, Y. (2008)
A non-stationary integer-valued autoregressive model.
\textit{Statistical Papers} \textbf{49}, 485--502.

\bibitem[Koopman et al.(2017)]{koopman17}
Koopman, S.J., Lit, R., Lucas, A. (2017)
Intraday stochastic volatility in discrete price changes: The dynamic skellam model.
\textit{Journal of the American Statistical Association} \textbf{112}, 1490--1503.

\bibitem[Liu et al.(2022)]{liu22}
Liu, M., Zhu, F., Zhu, K. (2022)
Modeling normalcy-dominant ordinal time series: An application to air quality level.
\textit{Journal of Time Series Analysis} \textbf{43}, 460--478.

\bibitem[Maddala(1983)]{maddala83}
Maddala, G.S. (1983)
\textit{Limited-Dependent and Qualitative Variables in Econometrics}.
Cambridge University Press, Cambridge.

\bibitem[McKenzie(1985)]{kenzie85}
McKenzie, E. (1985)
Some simple models for discrete variate time series.
\textit{Water Resources Bulletin} \textbf{21}, 645--650.

\bibitem[Michel \& de Jong(2018)]{michel18}
Michel, J., de Jong, R.M. (2018)
Mixing properties of the dynamic Tobit model with mixing errors.
\textit{Economics Letters} \textbf{162}, 112--115.

\bibitem[Newey \&  McFadden(1994)]{newey94}
Newey, W.K.,  McFadden, D. (1994)
Large sample estimation and hypothesis testing.
In R.F.\ Engle and D.\ MacFadden (eds): \textit{Handbook of Econometrics}, Vol.\ \textbf{4}, North-Holland, Amsterdam, 2111--2245.

\bibitem[O'Donovan(1983)]{odonovan83}
O'Donovan, T.M. (1983)
\textit{Short Term Forecasting: An Introduction to the Box--Jenkins Approach}.
John Wiley \& Sons, Chichester.

\bibitem[Powell(1984)]{powell84}
Powell, J.L. (1984)
Least absolute deviations estimation for the censored regression model.
\textit{Journal of Econometrics} \textbf{25}, 303--325.

\bibitem[Risti\'c et al.(2016)]{ristic16}
Risti\'c, M.M., Wei\ss{}, C.H., Janji\'c, A.D. (2016)
A binomial integer-valued ARCH model.
\textit{International Journal of Biostatistics} \textbf{12}, 20150051.

\bibitem[Tobin(1958)]{tobin58}
Tobin, J. (1958)
Estimation of relationships for limited dependent variables.
\textit{Econometrica} \textbf{26}, 24--36.

\bibitem[Wei{\ss}(2009)]{weiss09}
Wei{\ss}, C.H. (2009)
Modelling time series of counts with overdispersion.
\textit{Statistical Methods and Applications} \textbf{18}, 507--519.

\bibitem[Wei{\ss}(2018)]{weiss18}
Wei{\ss}, C.H. (2018)
\textit{An Introduction to Discrete-valued Time Series}.
John Wiley \& Sons, Chichester.

\bibitem[Wei{\ss} \& Aleksandrov(2022)]{weissaleksandrov22}
Wei{\ss}, C.H., Aleksandrov, B. (2022)
Computing (bivariate) Poisson moments using Stein--Chen identities.
\textit{The American Statistician} \textbf{76}, 10--15.

\bibitem[Wei{\ss} et al.(2022)]{weissetal22}
Wei{\ss}, C.H., Zhu, F., Hoshiyar, A. (2022)
Softplus INGARCH models.
\textit{Statistica Sinica} \textbf{32}, 1099--1120.

\bibitem[Xu et al.(2012)]{xu12}
Xu, H.-Y., Xie, M., Goh, T.N., Fu, X. (2012)
A model for integer-valued time series with conditional overdispersion.
\textit{Computational Statistics and Data Analysis} \textbf{56}, 4229--4242.

\bibitem[Zhu(2012)]{zhu12}
Zhu, F. (2012)
Modeling overdispersed or underdispersed count data with generalized Poisson integer-valued GARCH models.
\textit{Journal of Mathematical Analysis and Applications} \textbf{389}, 58--71.

\end{thebibliography}
\end{document}